\documentclass[preprint,authoryear]{elsarticle}

\usepackage[a4paper,margin=1in]{geometry}
\usepackage{fancyhdr}
\usepackage{dsfont}
\usepackage[utf8]{inputenc}
\usepackage[fleqn]{amsmath}
\usepackage{amssymb}
\usepackage{booktabs}
\usepackage{longtable}
\usepackage{xcolor}
\usepackage{appendix}
\usepackage{stmaryrd}
\usepackage{subfig}
\usepackage{hyperref}
\usepackage{graphicx}

\graphicspath{ {./} }

\usepackage{float}
\usepackage{layout} 
\title{The role of viscous regularization in dynamical problems, strain localization and mesh dependency }
\journal{Computer Methods in Applied Mechanics and Engineering}

\begin{document}

\author[1]{Alexandros Stathas}
\author[1]{Ioannis Stefanou\corref{cor1}}
\ead{ioannis.stefanou@ec-nantes.fr}
\cortext[cor1]{Corresponding author}
\address[1]{Institut de Recherche en Génie Civil et Mécanique (UMR CNRS 6183), Ecole Centrale de Nantes, Nantes, France}

\begin{abstract}
\noindent Strain localization is responsible for mesh dependence in numerical analyses concerning a vast variety of fields such as solid mechanics,  dynamics, biomechanics and  geomechanics. Therefore, numerical methods that regularize strain localization are paramount in the analysis and design of engineering products and systems. In this paper we revisit the elasto-viscoplastic, strain-softening, strain-rate hardening model as a means to avoid strain localization on a mathematical plane in the case of a Cauchy continuum. Going beyond previous works (\cite{DeBorst2020,needleman1988material,
Sluys1992,wang1997viscoplasticity}), we assume that both the frequency $\omega$ and the wave number $k$ belong to the complex plane. Therefore, a different expression for the dispersion relation is derived. We prove then that under these conditions strain localization on a mathematical plane is possible. The above theoretical results are corroborated by extensive numerical analyses, where the total strain and plastic strain rate profiles exhibit mesh dependent behavior.   
\end{abstract}

\maketitle

\section{Introduction \label{sec: one}}

\noindent Strain localization is a phenomenon which is found throughout natural and man-made structures. 
It is characterized by nonlinearity as well as different characteristic time and spatial scales ranging from the near instantaneous fracture of brittle materials to the vast geological time required for the formation of intricate patterns in the earth strata. Strain localization is responsible for mesh dependence in numerical analyses involving the Cauchy continuum, which is observed in a vast variety of applications in solid mechanics, dynamics, biomechanics, geomechanics and rock mechanics. Therefore, numerical techniques that correctly regularize strain localization are of great importance in the analysis and design of engineering products and systems.\\
\newline\noindent On a mathematical level, strain localization is understood as a bifurcation from the initial homogeneous deformation state of the structure to another equilibrium path. This automatically raises questions concerning the uniqueness of the reference homogeneous solution and its stability. 
The question of uniqueness of solution is decided from the conditions needed such that the determinant of the acoustic characteristic tensor of the problem is equal to zero \cite{Rice1976new,Rudnicki1975}. As far as the evaluation of stability is concerned, different criteria with varying degrees of applicability have been proposed in the literature (see \cite{bigoni2012nonlinear,chambon2004loss}).
\\
\newline\noindent In this paper the question of stability of the obtained equilibrium paths is decided based on the Lyapunov analysis \cite{chambon2004loss,lemaitre2020mecanique,
mawhin2005alexandr,Rice1976new,stefanou2016fundamentals}.  In a Classical Cauchy continuum it has been proven that in the context of a strain softening plasticity yield criterion, localization happens on a mathematical plane. This renders the solution obtained from numerical methods, such as the Finite Element method, mesh dependent. 
However, experimental evidence in materials suggests that localization in nature does not occur on a mathematical plane, rather it involves a small zone of finite thickness that accommodates the majority of the deformation (see \cite{chambon2004loss,Chester1998,Muhihaus1988,Sibson2003,sulem1995} among others). To remedy this inconsistency between experiments and analytical and numerical predictions two main approaches are often found in the literature. The first approach seeks to incorporate a modified constitutive law including the effects of viscosity \cite{DeBorst2020,Sluys1992,Sluys1993,wang1996interaction,
wang1997viscoplasticity}, while the other starts from the introduction of micromorphic continua, 
which introduce characteristic length scales to the mathematical problem \cite{DeBorst1991a,DeBorst1991,germain1973method,Rattez2018a,Sluys1993,
sulem2011stability,vardoulakis2009lecture}.\\
\newline \noindent Going beyond and expanding on existing results, we revisit, in this paper, the role of viscosity for the regularization of the classical Cauchy continuum. Considerable amount of research has been made on viscous regularization of localization under both quasi-static and dynamic conditions (see \cite{Loret1990,needleman1988material,Sluys1992}). In particular, for the quasi-static case, it is noted that the elasto-viscoplastic Cauchy medium based on a power law for the viscoplasticity description, will exhibit strain localization on a mathematical plane except if a particular procedure for the time integration takes place \cite{needleman1988material}. Furthermore, it is mentioned in \cite{Sluys1992} that the regularizing properties of the elasto-viscoplastic medium are present only in the context of dynamical analyses, due to the regularizing role of the higher order inertial terms, which are naturally introduced.  Moreover, for the dynamic case, a conclusion taken from \cite{needleman1988material,Sluys1992} is that the selection of a consistency or Perzyna yield condition for the viscoplastic model is prefered over a power law based on the fact that the third order terms of the Partial Differential Equation (PDE), gradually vanish in the latter as strain softening occurs. As it has been already presented in \cite{Abellan2006} and in particular in \cite{DeBorst2020} the formulation of the elasto-viscoplastic material model, specifically the position of the damper element into the idealized viscoplastic configuration, is of great importance. It has been shown in \cite{DeBorst2020} that a solution localizes into a mathematical plane once the configuration of the viscosity dashpot, the plasticity element and the elastic spring are in series. The present paper addresses stability and localization questions for the parallel configuration as studies  are not conclusive yet. Hence the elasto-viscoplastic model  with strain softening, which involves a parallel connection between the plasticity element and the dashpot, is examined at present using bifurcation and Lyapunov stability analysis. In particular, we examine in detail the stability of the reference solution of uniform deformation of the elasto-viscoplastic problem.\\
\\\noindent In previous works (\cite{DeBorst2020,needleman1988material,Sluys1992,wang1997viscoplasticity}), the question of localization of the deformation was addressed looking at the propagation velocity of localization. In particular it is mentioned in \cite{Sluys1992} that the original ill-posed problem of strain-softening plasticity presents imaginary wave speeds corresponding to standing waves, which cannot extend the localization zone. This is thought to be remedied by the introduction of viscosity. It is also mentioned that to properly test the conditions under which strain localization is present in the non-linear elasto-viscoplastic problem, we would require a closed form analytical expression for the solution, which until now is impossible. Since no analytical known way exists in solving the softening, rate-dependent plasticity problem, the focus was shifted in the derivation of the dispersion relationships. Namely, in the previous, it was assumed that if every mode has a real velocity, then the corresponding part of the deformation will propagate and therefore, it will not concentrate in only one element of the model as it would be the case with a standing wave.
An additional argument that is presented in \cite{Sluys1992,wang1996interaction}, is the dispersive character of the partial differential equation, which is assumed to further regularize the problem as the deformation front widens due to the different velocity of the deformation modes. In all these works, the dispersion relation was derived based on the assumption that the circular frequency $\omega$
and ,therefore, the wave velocities $c$ are real, while the wavenumber $k$ 
is complex: $k=k_r+k_i i,\;k_r,k_i=\alpha \in \mathbb{R}$, indicating spatial attenuation of the derived deformation modes, thus leading to characteristic localization length $l=\frac{1}{\alpha}$. However, we find no reason to strictly assume $\omega \; \in \;\mathbb{R}$, which makes an important difference in the analysis. The concept of imaginary  frequency can be shown to correspond to the real and negative square of the wave velocity $c$, which leads to divergence growth according to \cite{Rice1976new} (see also equation \eqref{ubar_expanded} in section \ref{sec: Bifurcation_analysis}). In a physical context imaginary frequencies $\omega$ are important in the description of physical phenomena as shown in the kinematic theory developed by \cite{hayes1970kinematic} and \cite{poeverlein1962sommerfeld}. \\
\\\noindent In this paper we depart from this main assumption by assuming both $\omega,k \in \mathbb{C}$, thus considering the problem in its general form. Furthermore, due to application of the Lyapunov analysis we are interested in the magnitude of the imaginary part of $\omega, \; \omega_i=\text{Im}[\omega]$, which controls the evolution of the amplitude of the oscillations and therefore, the stability of the reference homogeneous deformation state. We argue that if the amplitude of the mode with zero wavelength 
increases faster than that of the others, then this mode will dominate the deformation profile and strain localization on the mathematical plane 
will be possible. Theoretical proofs are presented showing this phenomenon and also, the emergence of traveling waves of strain localization in some cases. The theoretical results are corroborated by numerical analyses where the total strain and plastic strain profiles exhibit mesh dependent behavior. These analyses provide counter examples to viscous regularization in dynamical problems. \\
\newline\noindent This paper is organized as follows: In section \ref{sec: Bifurcation_analysis} the linearized differential equation describing the 1D elasto-viscoplastic model equiped with a von-Mises viscoplasticity consistency criterion is presented based on the works of \cite{DeBorst2020,Sluys1992,Wang1996,wang1996interaction,wang1997viscoplasticity}. In section \ref{sec: three} the dispersion relation of the linearized equation is derived between the wavenumber $k$ and the frequency $\omega$, both assumed to belong in the complex plane $\mathbb{C}$. Assuming $k$ to be complex is indicative of waves exhibiting attenuation or amplification as they travel through space, fairly common in a variety of applications. The assumption of a complex $\omega$ is crucial in order for the Lyapunov exponent $s=-i\omega$ to acquire negative or  positive real part, which controls the stability of the reference homogeneous deformation state. Applications exhibiting waves with complex $\omega$ can be found in \cite{Bernard2001,Deschamps1997,Gerasik2010,mainardi1984linear,mainardi1987energy,marion2013classical}. The introduction of the Lyapunov exponent $s$ in the context of the dispersion analysis is a fundamental difference of our work compared to previous attempts. The dispersion relation provides a link between the Lyapunov exponent that controls the evolution of the bifurcation and the wave number, which characterizes the localization width. Thus we can specify which bifurcation evolves the fastest determining the width of localization. We establish that the localization width $\lambda$ tends to zero (localization on a mathematical plane) for values of $\omega$ at the poles of the dispersion function $k(\omega)$. Furthermore, we present the concept of localization for complex $\omega,k$.
In section \ref{sec: Numerical_Analysis} we present a series of numerical analyses that corroborate the above theoretical findings. 
\section{The elasto-viscoplastic wave equation}\label{sec: Bifurcation_analysis}
\subsection{Problem description}
\noindent Let us consider a body under homogeneous small deformation lying at rest. The equilibrium equation is given as:
\begin{flalign}
&\sigma^\star_{ij,j}=0,
\end{flalign}
where $\sigma^\star_{ij}$ is the developed stress field, $i,j$ are indices indicating the directions of the components of the stress field $(i,j=1...3)$. In the above relation ($,j$) indicates the partial spatial derivative with respect to the $j$ direction. The Einstein summation condition over repeated indices is implied.
Considering a perturbation $\tilde{u}_i$  to the reference displacement field $u^{\star}_i$ of homogeneous deformation, we find a relationship between the perturbed stress and displacement $\tilde{u}$ fields, according to the conservation of linear momentum:
\begin{flalign}
&(\sigma^\star_{ij}+\tilde{\sigma}_{ij})_{,j}=\rho\ddot{\tilde{u}}_{i},\\
&\tilde{\sigma}_{ij,j}=\rho\ddot{\tilde{u}}_{i},\label{perturbed_equation}
\end{flalign}
where $(\cdot)$ corresponds to the time derivative and $\rho$ is the density of the material.
Both displacement and stress variations are arbitrary respecting only the boundary and loading conditions such that $\tilde{u}_i=0,\tilde{\sigma}_{i,j}n_j=0$ at the boundary of the body, where displacement and loading conditions are specified, respectively. In order to continue with the bifurcation analysis of the problem we need to look first at the 
elasto-viscoplastic constitutive law we take into account.
\subsection{Elasto-viscoplastic constitutive relations}

As mentioned in section \ref{sec: one} a variety of yield criteria and flow rules are available for modeling viscoplasticity. Here the von Mises yiled criterion with strain-hardening (softening) and strain-rate hardening is used together with the consistency approach \cite{wang1997viscoplasticity}.

\subsubsection{von Mises yield criterion consistency approach}
In an elasto-viscoplastic formulation the following relations hold as given in \cite{DeBorst1991a,Wang1996,wang1996interaction,wang1997viscoplasticity}:
\begin{flalign}
&F(\sigma_{ij},\bar{\epsilon}^{vp},\dot{\bar{\epsilon}}^{vp})=0,\\
&\dot{\varepsilon}_{ij}=\dot{\varepsilon}^e_{ij}+\dot{\varepsilon}^{vp}_{ij},\\
&\dot{\sigma}_{ij}=M^e_{ijkl}\left(\dot{\varepsilon}_{kl}-\dot{\varepsilon}^{vp}_{kl}\right)\label{total_stress_main},\\
&\dot{\varepsilon}^{vp}_{ij}=\dot{\lambda}\frac{\partial F}{\partial \sigma_{ij}}\label{varepsilon_vp_main},\\
&\bar{\epsilon}^{vp} = \int^t_0\dot{\bar{\epsilon}}^{vp}dt,
\end{flalign}
where $F=F(\sigma_{ij},\bar{\epsilon}^{vp},\dot{\bar{\epsilon}}^{vp})$ is the yield function incorporating the effects of strain and strain-rate hardening through the use of the accumulated deviatoric viscoplastic strain $\bar{\epsilon}^{vp}$ and its rate $\dot{\bar{\epsilon}}^{vp}$, respectively.
The viscoplastic multiplier $\dot{\lambda}$ is given by the consistency condition: $\dot{F}=0,\;\dot{\lambda}F=0$. The time derivative of the yield condition in this case is the following:
\begin{flalign}
\dot{F}= \frac{\partial F}{\partial \sigma_{ij}}\dot{\sigma}_{ij}+\frac{\partial F}{\partial \bar{\epsilon}^{vp}}\dot{\bar{\epsilon}}^{vp}+\frac{\partial F}{\partial \dot{\bar{\epsilon}}^{vp}}\ddot{\bar{\epsilon}}^{vp}=0.
\end{flalign}
The von Mises yield criterion with strain-hardening (softening) and strain-rate hardening for the consistency approach reads \cite{Wang1996,wang1996interaction,wang1997viscoplasticity}:\\
\begin{flalign}
F(\sigma_{ij},\bar{\varepsilon}_{ij},\dot{\bar{\varepsilon}}_{ij})=\sqrt{3J_2(\sigma_{ij})}-(F_0+h\bar{\epsilon}^{vp}+g\dot{\bar{\epsilon}}^{vp})\label{Von_Mises_yield_Criterion_main},
\end{flalign}
where $F_0$ is the initial yield strength of the material, $h$ is a parameter indicating strain hardening of the material ($h<0$ indicates strain softening) with increasing accumulated plastic strain and $g$ is a parameter indicating hardening of the material with increasing plastic strain-rate ($g<0$ indicates strain-rate softening). The accumulated plastic strain is defined as $\dot{\bar{\epsilon}}^{vp} = \sqrt{\frac{2}{3}\dot{\varepsilon}^{vp}_{ij}\dot{\varepsilon}^{vp}_{ij}}$. Only strain rate hardening ($g>0$) is examined here.\\
\newline \noindent Alternatively one can define the classic von Mises yield condition (without the term $g\dot{\bar{\epsilon}}^{vp}$) and assume that after the material reaches the yield limit the viscoplastic strain-rate is given as $\dot{\varepsilon}^{vp}_{ij} =\frac{F}{\eta F_0}\frac{\partial F}{\partial \sigma_{ij}}$. This model is known as the Perzyna model. The results are the same as far as monotonic loading is applied and non-holonomic behaviour of the material is excluded, provided that $g=\eta F_0$. In the case of stress reversal and subsequent unloading, however, the results between the consistency approach and the Perzyna model will be different due to the elasto-viscoplastic component that the Perzyna model predicts during unloading (see \cite{Heeres2002}). \\
\newline\noindent In what follows the principal results of the elasto-viscoplastic bifurcation analysis are presented. A full derivation of the elasto-viscoplastic constitutive relations is presented in the Appendix \ref{Appendix: A}. Applying equation \eqref{Von_Mises_yield_Criterion_main}
to equations \eqref{total_stress_main}  and \eqref{ldot} of the Appendix \ref{Appendix: A}, we arrive  at the relationship for the stress rate $\tilde{\sigma}_{ij}$:
\begin{flalign}
&\tilde{\sigma}_{ij}=M^e_{ijkl}\left(\tilde{\varepsilon}_{kl}-\frac{C}{-h+C}\tilde{\varepsilon}_{kl}+\frac{h b_{kl}}{-h+C}\dot{\tilde{\lambda}}\right)\label{constitutive_law_perturbed_main},\\
\intertext{where in order to simplify the notation we have replaced accordingly:}
&\frac{\partial F}{\partial \dot{\lambda}}=h,\nonumber\\
&C=\frac{\partial F}{\partial \sigma_{ij}}M^e_{ijkl}\frac{\partial F}{\partial \sigma_{kl}},\nonumber\\
&b_{kl} = \frac{\partial F}{\partial \dot{\lambda}}\frac{\partial F}{\partial \sigma_{kl}}\label{substitution_rules}.
\end{flalign}
\subsection{Derivation of the perturbed equation\label{sec:perturbed_eq}}
\noindent We proceed now in  deriving the general linearized perturbed equation of equilibrium for the given material law under monotonic loading.
Inserting equation \eqref{constitutive_law_perturbed_main} into equation \eqref{perturbed_equation} and taking into account the spatial derivative of equation \eqref{varepsilon_vp_main}, we obtain:
\begin{flalign}
&M^e_{ijkl}\left(\tilde{\varepsilon}_{kl,j}-\frac{C}{-h+C}\tilde{\varepsilon}_{kl,j}+\frac{h}{-h+C}\dot{\tilde{\varepsilon}}_{kl,j}-{M^{e^{{-1}}}_{ijkl}}\rho\dddot{\tilde{u}}_{i}\right)=\rho \ddot{\tilde{u}}_{i},
\end{flalign}

\noindent This equation describes the spatio-temporal evolution of perturbations from the reference solution of homogeneous deformation in 3D.
\subsubsection{Shearing of a viscous Cauchy layer}
\noindent We constrain our analysis to the study of 1D problems, since they constitute the simplest case to study localization and the regularization effects coming from the above material law. In this way direct parallels can be drawn between our work and the main bulk of literature on the subject \cite{DeBorst2020,needleman1988material,Sluys1992,Wang1996}. 
For the shearing of an 1D layer we assume that the shearing is coaxial to the direction of $x_1$ and that the body deforms in the direction $x_1$. Therefore $\tilde{u}_i=[0,\tilde{u}_2]^T=[0,\tilde{u}]^T$. Since we are in a state of 1D deformation, only the derivatives along the 1D axis, $x_1$, survive, therefore we set $\frac{\partial \tilde{u}_2}{\partial x_1}=\frac{\partial \tilde{u}}{\partial x} $. Taking into account the appropriate material constant $M^e_{ijkl}=D^e_{2121}=G$, we proceed in deriving the perturbed linear momentum equation for the shearing of an 1D elasto-viscoplastic layer:
\begin{flalign}
&G\bar{h}\frac{\partial^2 \tilde{u}}{\partial x^2}-\frac{\partial^2 \tilde{u}}{\partial t^2}\frac{(3+\bar{h})G}{v^2_s}+\bar{\eta}^{vp}G\left(\frac{\partial^3 \tilde{u}}{\partial t \partial x^2}-\frac{1}{v^2_s}\frac{\partial^3 \tilde{u}}{\partial t^3}\right)=0\label{viscoelastic_wave_main},
\end{flalign}
where $v_s=\sqrt{\frac{G}{\rho}}\;\text{and}\;\bar{\eta}^{vp}G=\eta F_0=g$.\\
This coincides with the elasto-viscoplastic equation derived by \cite{DeBorst2020,Sluys1992,Wang1996,wang1997viscoplasticity}. However, the equation derived above describes the evolution of a perturbation from the initial homogeneous deformation state. It will not be used as a description of the total behavior of the material as it neglects the material behavior in unloading and we are not interested in the solution of the elasto-plastic problem but only at the stability of the homogeneous deformation state, in order to draw conclusions about strain localization \cite{lemaitre2020mecanique}. Here we note that the equation \eqref{viscoelastic_wave_main} has time independent coefficients (autonomous system, see \cite{Brauer1969}). 
\\
\newline \noindent As mentioned in \cite{wang1997viscoplasticity}, this equation contains two components, a classical elastoplastic wave equation plus the higher order rate-dependent terms. The nature of this differential equation is defined by the higher order derivatives. It is also stated in \cite{Sluys1992} that, in the limit of high viscosity $\bar{\eta}^{vp}\rightarrow \infty$, only the rate terms contribute, since they travel with the elastic wave velocity. In this case, the implied deformation pulse will travel with the corresponding elastic wave velocity as predicted in \cite{Loret1990} and \cite{needleman1988material}.
\subsubsection{Normalizing the 1D elasto-viscoplastic wave equation.}
\noindent We consider $\bar{u}=\frac{u}{u_c},\;\bar{t}=\frac{t}{t_c},\;\bar{x}=\frac{x}{x_c}$, where $u_c,\;t_c,\;x_c$ are the characteristic displacement, time and length, respectively. Applying these definitions  to equation
\eqref{viscoelastic_wave_main} we obtain:
\begin{flalign}
&\left(\frac{x^2_c}{v^2_s t^2_c}\frac{\partial^3\bar{u}}{\partial \bar{t}^3}-\frac{\partial^3\bar{u}}{\partial \bar{x}^2\partial \bar{t}}\right)\frac{\bar{\eta}^{vp}}{t_c \bar{h}}+\frac{x^2_c}{v^2_s t^2_c}\frac{3+\bar{h}}{\bar{h}}\frac{\partial^2 \bar{u}}{\partial \bar{t}^2}-\frac{\partial^2 \bar{u}}{\partial \bar{x}^2}=0.\\
\intertext{Introducing the characteristic velocity $v_c=\frac{x_c}{t_c}$, the result is written as:}
&\left(\frac{v^2_c}{v^2_s }\frac{\partial^3\bar{u}}{\partial \bar{t}^3}-\frac{\partial^3\bar{u}}{\partial \bar{x}^2\partial \bar{t}}\right)\frac{\bar{\eta}^{vp}}{t_c \bar{h}}+\frac{v^2_c}{v^2_s}\frac{3+\bar{h}}{\bar{h}}\frac{\partial^2 \bar{u}}{\partial \bar{t}^2}-\frac{\partial^2 \bar{u}}{\partial \bar{x}^2}=0\label{normalized equation_main}.
\end{flalign}
The above equation is linear and has solutions of the form:
\begin{flalign}
\bar{u}(\bar{x},\bar{t}) = A\exp{[i( \bar{k} \bar{x}-\bar{\omega} \bar{t})]}\label{u_bar},
\end{flalign}
where $\bar{k},\bar{\omega} \;\in \; \mathds{C}$ and $A\;\in\; \mathds{R}$ is a constant indicating the wave amplitude. Finally, inserting the non-dimensional solution \eqref{u_bar} into the normalized equation \eqref{normalized equation_main}  we arrive at:
\begin{flalign}
&\bar{k}^2 v^2_s (\bar{h} \bar{t}_c - i\bar{\eta}^{vp} \bar{\omega}) - v^2_c \bar{\omega}^2 [(3 + \bar{h}) \bar{t}_c - i\bar{\eta}^{vp}\bar{\omega}]=0\label{dispersion_relation_main}.
\end{flalign}
It is worth emphasizing that the choice of assuming both $\bar{\omega},\bar{k}\;\in \; \mathds{C}$ is not studied extensively in the literature. However, examples of the notion can be found in \cite{mainardi1984linear,mainardi1987energy,marion2013classical}.\\

\noindent Replacing $V=\frac{v_c}{v_s},\; T=\frac{\bar{\eta}^{vp}}{t_c}$ and solving the above for $\bar{k}$ we obtain:
\begin{flalign}
&\bar{k}_{1,2} =\pm \sqrt{\frac{V^2 \bar{\omega}^2 (3 + \bar{h} - i T \bar{\omega})}{(\bar{h} - i T \bar{\omega})}}\label{k_solution}.
\end{flalign}
The above equation can be also derived directly by use of a Fourier transform on equation \eqref{normalized equation_main}.
Expanding $\bar{k},\bar{\omega}$ in imaginary and real parts, $\bar{k}=\bar{k}_r+\bar{k}_ii,\; \bar{\omega}=\bar{\omega}_r+\bar{\omega}_i i$ as explained in \cite{hayes1970kinematic,Maugin2007a,poeverlein1962sommerfeld} indicates that in our analysis the dependence of the amplitude of the solution has both a spatial and a temporal component. In particular the non-dimensional solution can be written as:
\begin{flalign}
\bar{u}(\bar{x},\bar{t})=\exp{(-\bar{k}_i\bar{x}+\bar{\omega}_i\bar{t})}\exp[{i(\bar{k}_r\bar{x}-\bar{\omega}_r\bar{t})]}\label{ubar_expanded},
\end{flalign} 
where, without loss of generality, the amplitude constant in front of the exponential terms of solution \eqref{ubar_expanded} is set to unity. The first factor in the right hand side of equation \eqref{ubar_expanded} indicates a quantity that increases or decreases based on the relationship between $(\bar{\omega}_i\bar{t}$ and $\bar{k}_i\bar{x})$. In this paper, we define that an observer moving along $\bar{x}$ with a velocity $c_i$, such that the amplitude profile $\exp{(-\bar{k}_i\bar{x}+\bar{\omega_i} \bar{t})}$ remains constant, is moving with the \textit{amplitude velocity}:
\begin{flalign}
c_i = \frac{\bar{\omega_i}}{\bar{k}_i}\label{amplitude_velocity}.
\end{flalign}
Conversely the second factor of equation \eqref{ubar_expanded} indicates the classical wave solution. An observer moving with a velocity $c_r$ such that the phase $\exp{[i(\bar{k}_r \bar{x}-\bar{\omega}_r\bar{t})]}$ remains constant is said to be moving with the \textit{phase velocity} \cite{marion2013classical,Pain1993,Sluys1992}:
\begin{flalign}
c_r = \frac{\bar{\omega_r}}{\bar{k}_r}\label{phase_velocity}.
\end{flalign}
According to the definition of Lyapunov for continuous dynamical systems \cite{Rattez2018b,mawhin2005alexandr}, for an equilibrium solution to be unstable, the amplitude of the initial perturbation must increase in time. According to Lyapunov stability analysis, a partial solution of the partial differential equation \eqref{normalized equation_main} is given by $\bar{u}(\bar{x},\bar{t}) = \exp{(\bar{s}\bar{t}+i\bar{k} \bar{x})}$,
where $\bar{s}$ is the Lyapunov exponent $\bar{s}=-i\bar{\omega}$.
From this we conclude that the important term whose sign determines the stability of the reference solution, corresponding to homogeneous deformation \eqref{u_bar}, is the imaginary part of $\bar{\omega}$, i.e. $\bar{\omega}_i$. 
Therefore, for the perturbation to grow in amplitude, the term $\exp{(-\bar{k}_i\bar{x}+\bar{\omega_i} \bar{t})}$ must be increasing as the wave travels. Localization on a mathematical plane will happen if we can find appropriate $\bar{\omega}_i,\bar{k}_i$ terms for the amplitude to be constantly increasing the fastest for the smallest possible wavelength $\bar{\lambda} \rightarrow 0$ $(\bar{k}_r=\frac{2\pi}{\bar{\lambda}} \rightarrow \infty)$. 

\section{Dispersion analysis \label{sec: three}}
\subsection{Solution of the dispersion equation}
\noindent Equation \eqref{k_solution} consists of two multivalued functions $\bar{k}_1(\bar{\omega}),\bar{k}_2(\bar{\omega})$ in the complex set $\bar{\omega}\; \in \;\mathbb{C}$. Introduction of branch cuts along selected points of unambiguous value is needed for their study on the values of their argument $\bar{\omega}$ \cite{arfken1999mathematical}.\\ 
\noindent Noticing the square powers of $V^2,\;\bar{\omega}^2$ inside the root, equation \eqref{k_solution} can be simplified  yielding:
\begin{flalign}
&\bar{k}_{1,2}(\bar{\omega}) =\pm V \left(\frac{3+\bar{h}}{\bar{h}}\right)^{\frac{1}{2}}\bar{\omega}\left(\frac{3 + \bar{h}}{ T}i+ \bar{\omega}\right)^{\frac{1}{2}}\left(\frac{\bar{h}}{ T}+ \bar{\omega}\right)^{-\frac{1}{2}}\label{final_form_k1}.
\end{flalign}

\begin{figure}[h]
    \centering
    \includegraphics[width=0.9\textwidth]{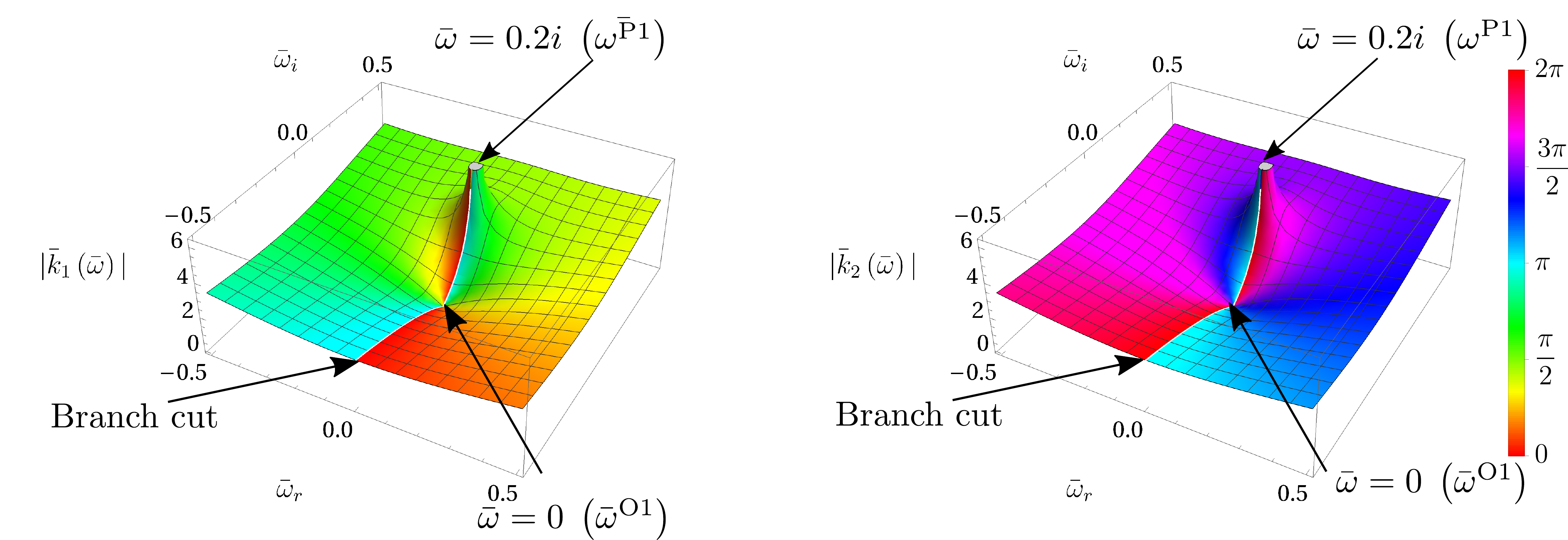}
    \caption{On the left: Complex 3D plot of the $\bar{k}_1(\bar{\omega})$ solution \eqref{k_solution}. On the right: Complex 3D plot of the $\bar{k}_2(\bar{\omega})$ solution \eqref{k_solution}.  Values on vertical axis indicate the solution's magnitude, where the coloring indicates the argument of the function. Along the branch cut discontinuity, the difference in color indicates the jump in the argument of $\bar{k}_1(\bar{\omega})$.}
    \label{fig:k_Complex3Dplot_1a_1b}
\end{figure}

\begin{figure}[h]
    \centering
    \includegraphics[width=0.9\textwidth]{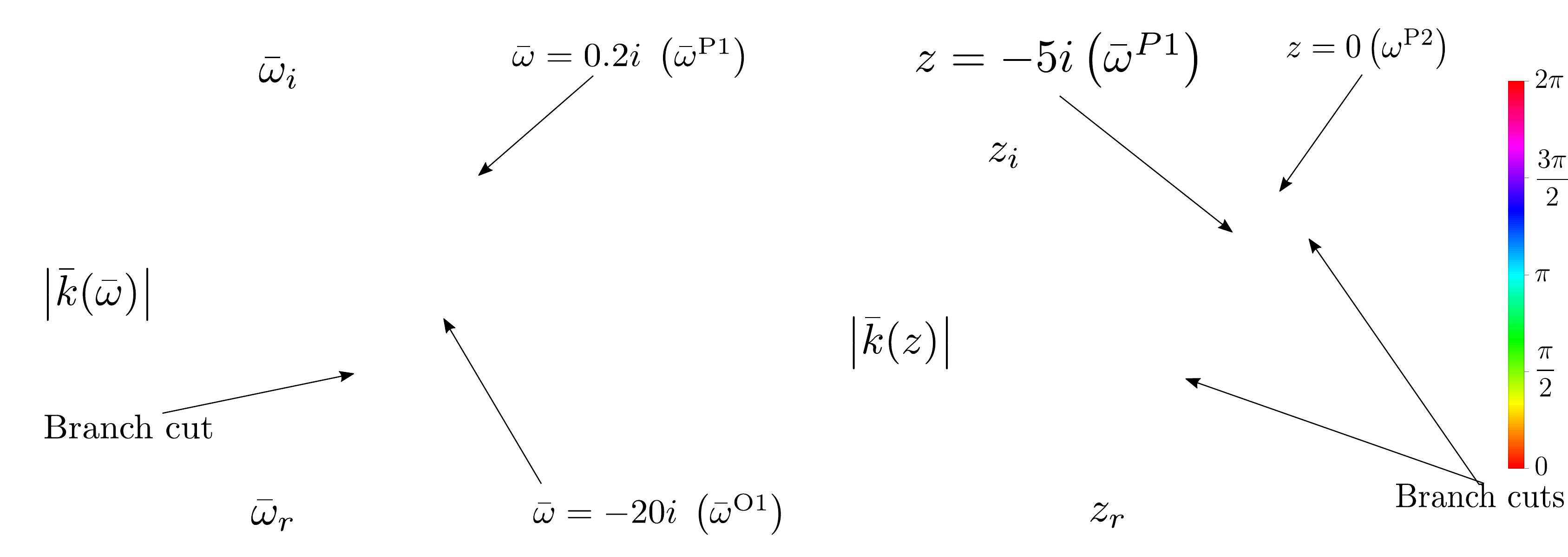}
    \caption{Complex 3D plot of the $\tilde{k}(\bar{\omega})$ combination of the two solutions. This envelope incorporates all the waves that travel to the positive part of the axis together with the highest spatial amplification coefficient. On the left: Complex 3D Plot of $\tilde{\bar{k}}(\bar{\omega})=|\bar{k}_r|+i|\bar{k}_i|$ for values of $\bar{\omega}_r,\;\bar{\omega}_i$ close to the pole value $\bar{\omega}^{\text{P1}}=0.2i$. On the right: Complex 3D Plot of $\tilde{\bar{k}}\left(\frac{1}{z}\right)=|\bar{k}_r|+i|\bar{k}_i|$ for values of $\bar{\omega}_r,\;\bar{\omega}_i$ close to infinity, when $z^{\text{P2}}=0,\; \left(\frac{1}{z}\right)\rightarrow \infty$ and the pole value $\bar{\omega}^{\text{P1}}=0.2i\rightarrow z^{P1}=-5i$}
    \label{fig: k_complex3D_plot}
\end{figure}

\noindent Each of the two solutions contains right and left propagating waves based on the sign combinations of $\bar{\omega}_r,\bar{k}_r$. The second solution $\bar{k}_2(\omega)$ presents the exact same points of interest as the first. The difference lies in the $-1$ factor between the two solutions. This factor according to the Euler identity can be written as $e^{i\pi}=-1$ and ,therefore, indicates a change in the argument of the second solution. Figure \ref{fig:k_Complex3Dplot_1a_1b} shows 3D plots of $\bar{k}_1(\bar{\omega}),\;\bar{k}_2(\bar{\omega})$. The colors on the right of Figure \ref{fig:k_Complex3Dplot_1a_1b} are changed indicating a change of the argument from the upper half of the imaginary plane to the lower half of it, meaning that the two solutions $\bar{k}_{1,2}(\bar{\omega})$ behave differently when it comes to the spatial amplification/attenuation coefficient $\bar{k}_i$. In particular, $\bar{k}_1(\bar{\omega})$ predicts only attenuation waves, while $\bar{k}_2(\bar{\omega})$ consists of amplification waves. This change indicates that the waves in Figure \ref{fig:k_Complex3Dplot_1a_1b} travel in the same direction with opposite imaginary part of $\bar{k}(\bar{\omega})$. To us the propagation direction of the wave is not important because of spatial symmetry of the solution \eqref{k_solution}.\\ 
\newline \noindent In the next section the points of interest of the $\bar{k}_1(\bar{\omega})$ are presented and their behavior is explained in the form of branch cuts and poles. As discussed previously, the same behavior is valid for $\bar{k}_2(\bar{\omega})$.
We choose to draw further conclusions in the form of plots over line from the combination of the two solutions $\bar{k}_{1,2}(\bar{\omega})$. In particular, we focus on the positive real parts of the solutions $\bar{k}_{1r}(\bar{\omega}),\bar{k}_{2r}(\bar{\omega})>0$ as well as the positive imaginary parts of the solutions $\bar{k}_{1i}(\bar{\omega}),\bar{k}_{2i}(\bar{\omega})>0$. Thus, 
we investigate the function $\bar{k}{(\omega)}=|\bar{k}_r(\bar{\omega})|+i |\bar{k}_i(\bar{\omega})| $ as shown later in Figure \ref{fig: k_complex3D_plot}.\\

\subsection{Poles and zeros \label{Poles_and_zeros}}
\noindent Studying equation \eqref{final_form_k1}, the following points can be readily specified in the above form:
\begin{itemize}
\item The third factor indicates a zero at the origin: $\bar{\omega}^{\text{O1}}=0$.
\item The fourth factor becomes zero at position: $\bar{\omega}^{\text{O2}}=-\frac{3+\bar{h}}{T}i$.
\item The last factor indicates the presence of a pole at: $\bar{\omega}^{\text{P1}}=-\frac{\bar{h}}{T}i$.
\item The value of the function at complex infinity $\bar{\omega}^{\text{P2}}\rightarrow\infty$ is found to be infinite in a complex sense, $\lim_{\bar{\omega}\rightarrow \bar{\omega}^{\text{P2}}} \bar{k}_1(\bar{\omega}) \rightarrow \infty$.
\end{itemize}
For the purposes of our analysis the behavior of the dispersion function at the poles $\bar{\omega}^{\text{P1}},\;\bar{\omega}^{\text{P2}}$ is very important as it will be shown to promote localization on a mathematical plane. 

\begin{figure}[h]
    \centering
    \includegraphics[width=0.9\textwidth]{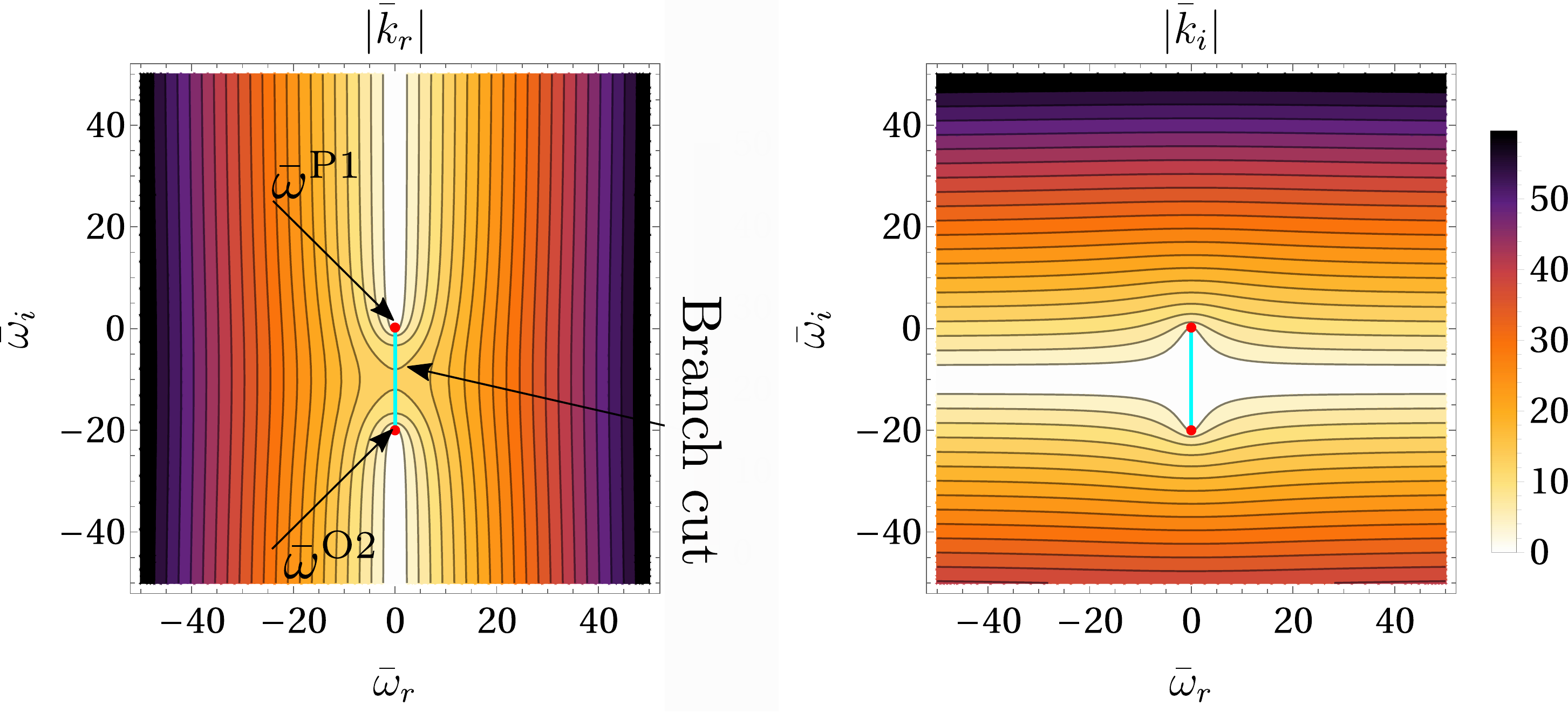}
    \caption{Contours of the solution \eqref{k_solution} indicating with red color the position of the branch points and with cyan the branch cut line that connects them.}
    \label{fig:branch_cut_plot}
\end{figure}

\noindent Because of the fractional powers of the second and third term, equation \eqref{final_form_k1} is a multivalued equation, since it is affected by the values of the argument. In order to remove the ambiguity from the function we need to constrain it in such a way that each value of the function corresponds to only one argument. For this we introduce \textit{branch cuts}. A branch cut is a discontinuity in the function that is defined by arbitrarily joining the two points defined as branch points. The branch points are defined as points of unambiguous value, where the argument of the function is exactly known for a particular value of the function and the values corresponding to other points in a region sufficiently close to the branch point depends on the argument of the complex number inserted in the function. Two such points for the complex function $f(\bar{\omega})=\bar{\omega}^{\frac{1}{2}}$ are $\bar{\omega}=0,\;\text{and}\;|\bar{\omega}|\rightarrow\infty,\;\bar{\omega} \in \mathbb{C}$, because for these particular numbers the value of the function is always zero and infinity respectively. However, around them the value of the function depends on the argument of the complex number (see below).\\
\newline\noindent We can translate this result to other points in the complex plane, namely to $\bar{\omega}^{\text{O2}},\;\bar{\omega}^{\text{P1}}$. In a region close and around $\bar{\omega}^{\text{O2}}=-i\frac{3+\bar{h}}{T}$ the complex number with starting point $\bar{\omega}^{\text{O2}}$ that follows the curve surrounding $\bar{\omega}^{\text{O2}}$ changes its argument by $2\pi i$. However, the factor $\left(\frac{3 + \bar{h}}{i T}- \bar{\omega}\right)^{\frac{1}{2}}$ only changes by $\pi i$, meaning that there is a sign difference between the starting and the end position along the closed curve at the same point. Similarly, the same happens in the region near the pole $\bar{\omega}^{\text{P1}}$, where for every $2\pi i$ that the relative complex number starting at $\bar{\omega}^{\text{P1}}$ changes following the surrounding curve, the factor $ \left(\frac{\bar{h}}{ i T}- \bar{\omega}\right)^{-\frac{1}{2}}$ changes by $-\pi i$. However at complex infinity $(\bar{\omega} \rightarrow \infty)$, both previous points are entailed by the curve at infinity. Therefore, the total change in the argument is $\pi i-\pi i=0$. This means that, $\bar{\omega}^{\text{P2}}$ it is not a branch point (it does not belong to the branch cut). The simplest cut is the one that follows the line defined by the two branch points as shown in Figure \ref{fig:branch_cut_plot}.
Since the point at complex infinity is not a branch point then it can be expected to be an isolated singular point, namely a pole of n-order or an essential singularity \cite{arfken1999mathematical,brown2009complex}. In this case it can be proven to be a simple pole as shown in Appendix \ref{Appendix_B}\\
\newline\noindent Introducing the mapping $\bar{\omega}=\frac{1}{z}$ we notice $\bar{\omega}\rightarrow\infty$ can be written as $\frac{1}{z}$ when $ z\rightarrow 0$. The properties of this mapping are explained in \cite{brown2009complex} and in the Appendix \ref{Appendix_B}.  

\qquad

\begin{figure}[h]
    \centering
    \includegraphics[width=0.9\textwidth]{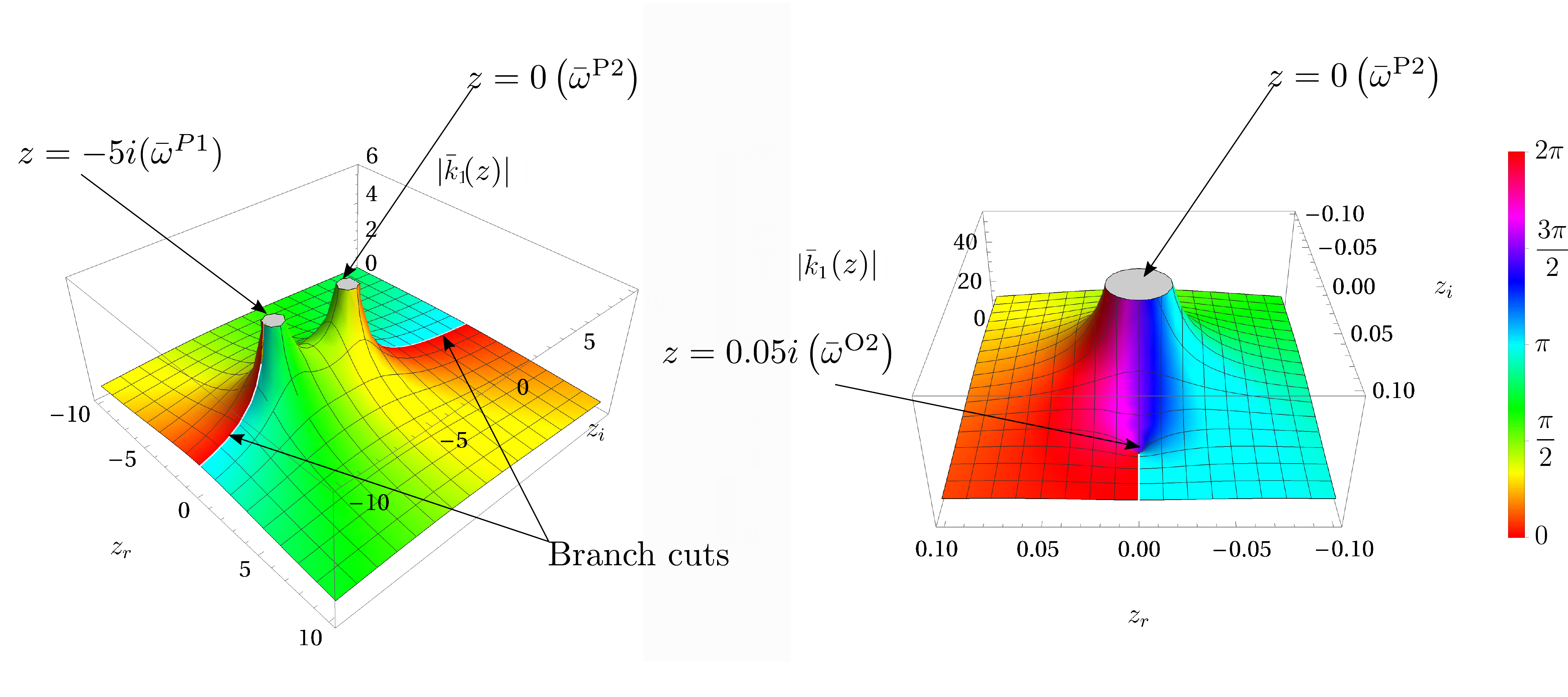}
    \caption{Complex 3D plot of the $\bar{k}_1(z)$ where $z=\frac{1}{\bar{\omega}}$. On the left part of the Figure the behavior in the region near $\bar{\omega}$ infinity $\bar{\omega}^{\text{P2}}$ and the pole at $\bar{\omega}^{\text{P1}}$ is presented.We notice the two poles lying at positions $\bar{\omega}^{\text{P2}}:\;z=0$ and $\bar{\omega}^{\text{P1}}$ lying at $\omega^{\text{P1}}: z^{p}=-5i$ respectively. On the right the region close to the pole at infinity $z=0$ is plotted. We notice the existence of the zero $\bar{\omega}^{\text{O2}}$ that due to the mapping now lies at $z=0.05$ extremely close to $z=0$, inside the unit circle.}
    \label{fig:k_Complex3Dplot_inf_pole}
\end{figure}\qquad

\noindent The plots showing the poles and zeros of the function $\bar{k}_1(z)$ with the mapping are shown in Figure \ref{fig:k_Complex3Dplot_inf_pole}. On the right part of the Figure the region around the poles and infinity is shown while on the left a detail is presented where the zero value $\bar{\omega}^\text{O2}$ - that due to the mapping is found closer to the origin - is shown.\\


\subsection{Localization on a mathematical plane}
\noindent Localization will happen when the amplitude of a particular perturbation mode as shown in equation \ref{ubar_expanded} is found to be continuously increasing faster than the rest. If this happens for the perturbation of the smallest possible wavelength $\bar{\lambda}\rightarrow 0$ that corresponds to $\bar{k}_r\rightarrow \infty$ then localiazation on a mathematical plane takes place. \\ 
\newline
\noindent As stated in the previous paragraphs without loss of generality we focus on the positive real and imaginary parts of the function $|\bar{k}({\bar{\omega}})|=|\bar{k}_r|+|\bar{k}_i|i$. $|\bar{k}(\bar{\omega})|$ has the same poles $\bar{\omega}^P$ and zeros $\bar{\omega}^O$ as the original $\bar{k}(\bar{\omega})$ with the added simplification that only the positive argument values of both functions 
$\bar{k}_1(\bar{\omega}),\bar{k}_2(\bar{\omega})$ are plotted. This simplifies our analysis with regards to the sign of $\bar{k}_i$, which contributes to the exponential growth of the amplitude, but it is not crucial for the time evolution of the perturbation which is determined by $\bar{\omega}$.\\ 
\newline \noindent First we focus to the pole value at $\bar{\omega}^{\text{P1}}$ which is shown in the left 3D plot of Figure \ref{fig: k_complex3D_plot}. There the pole $\bar{\omega}^{P1}$ and the first zero $\bar{\omega}^{O1}$ are shown. The pole lies at the value $\bar{\omega}^{\text{P1}}=-\frac{\bar{h}}{\bar{T}}i$ with $\bar{h}<0$, corresponding to a real and positive Lyapunov exponent $\bar{s}=i\bar{\omega}=\frac{\bar{h}}{\bar{T}}>0$. Since for $\bar{k}_r(\bar{\omega}) \rightarrow \infty$ when $\bar{\omega}=\bar{\omega}^{\text{P1}}$, we conclude that localization on a mathematical plane is possible $(\bar{\lambda}=\frac{2\pi}{\bar{k}_r}\rightarrow 0)$, while in this case the rate of increase of the perturbation amplitude is bounded.\\
\noindent The perturbation growth coefficient for viscoplastic media is bounded in contrast to rate-independent media where the perturbation growth coefficient is infinite when the conditions for strain localization are met $(\bar{s}\approx \frac{1}{\bar{\lambda}})$. However, in both cases strain localization happens on a mathematical plane and, therefore, this analysis shows that viscoplasticity does not regularize this problem even in the presence of inertia terms.
\newline
\noindent The behavior of $|\bar{k}(\bar{\omega})|$ at $\bar{\omega} \rightarrow \infty$ meaning $\bar{\omega}=\bar{\omega}^{P2}$ cannot be captured in this plot. 
For this reason we perform a change of variables in $\bar{\omega}$, replacing with $\bar{\omega}=\frac{1}{z}$. Now the value of $\bar{\omega}$ at infinity corresponds to $z=0$. The function $\bar{k}(z)$ is plotted on the right of  Figure \ref{fig: k_complex3D_plot}, where we capture the value at $\bar{\omega}^{\text{P2}}\rightarrow\infty$ when $z=0 \;(\bar{\omega}^{P2})$ and at the pole $\bar{\omega}^\text{P1}=\frac{1}{z^\text{P1}}$.The second zero value is also captured in the line plots of Figures \ref{fig: dispersion_kr_omegai} and \ref{fig: dispersion_ki_omegai} with the help of the transformation $\bar{k}(z)$. In this case $\bar{k}(z)$ tends to infinity. Therefore $\omega^{P2}$ also constitutes a localization point.\\ 

\subsection{Characterization of the waves in the complex plane \label{sec: Characterization of the waves in the complex plane}}
\noindent Having qualitatively described the behavior of the solutions of the dispersion equation we proceed with assigning specific values to the dimensionless parameters according to the material parameters taken from \cite{DeBorst2020}. These values are presented in Table \ref{Table: material_params} and describe the case of a viscoplastic material obeying the von Mises yield criterion with strain-hardening $(h>0)$ and strain-rate hardening $(g>0)$, M1, as well as the case of strain-softening $(h<0)$ and strain-rate hardening $(g>0)$, M2.\\ 
\begin{table}
\begin{center}
\begin{tabular}{ l l l} 
\hline
  non-dimensional  & M1  & M2\\
  material parameters & stable & unstable\\
 \hline
 $\bar{V}=\frac{v_c}{v_s}$ & 0.25 & 0.25 \\ 
 $\bar{T}=\frac{g}{t_c G}$ & 0.15 & 0.15 \\ 
 $\bar{h}=\frac{h}{G}$ & 0.03 & -0.03\\ 
 \hline
\end{tabular}
\caption{\label{Table: material_params}Non-dimensional material parameters used for the numerical analyses}
\end{center}
\end{table}

\noindent We focus our attention on the case M2. By standing waves here we refer to profiles stationary in space whose values, however, depend on time due to the exponential growth coefficient $\bar{\omega}_i$.  
The contours of the real $|\bar{k}_r|$ and imaginary parts $|\bar{k}_i|$ of the combination of solutions near the pole $\omega^{P1}$
are presented in Figure \ref{fig: dispersion_kr_ki}. We can define three cases for an 1D elasto-viscoplastic medium expanding to infinity in both directions around the origin:  

\begin{figure}[h]
    \centering
    \includegraphics[width=0.9\textwidth]{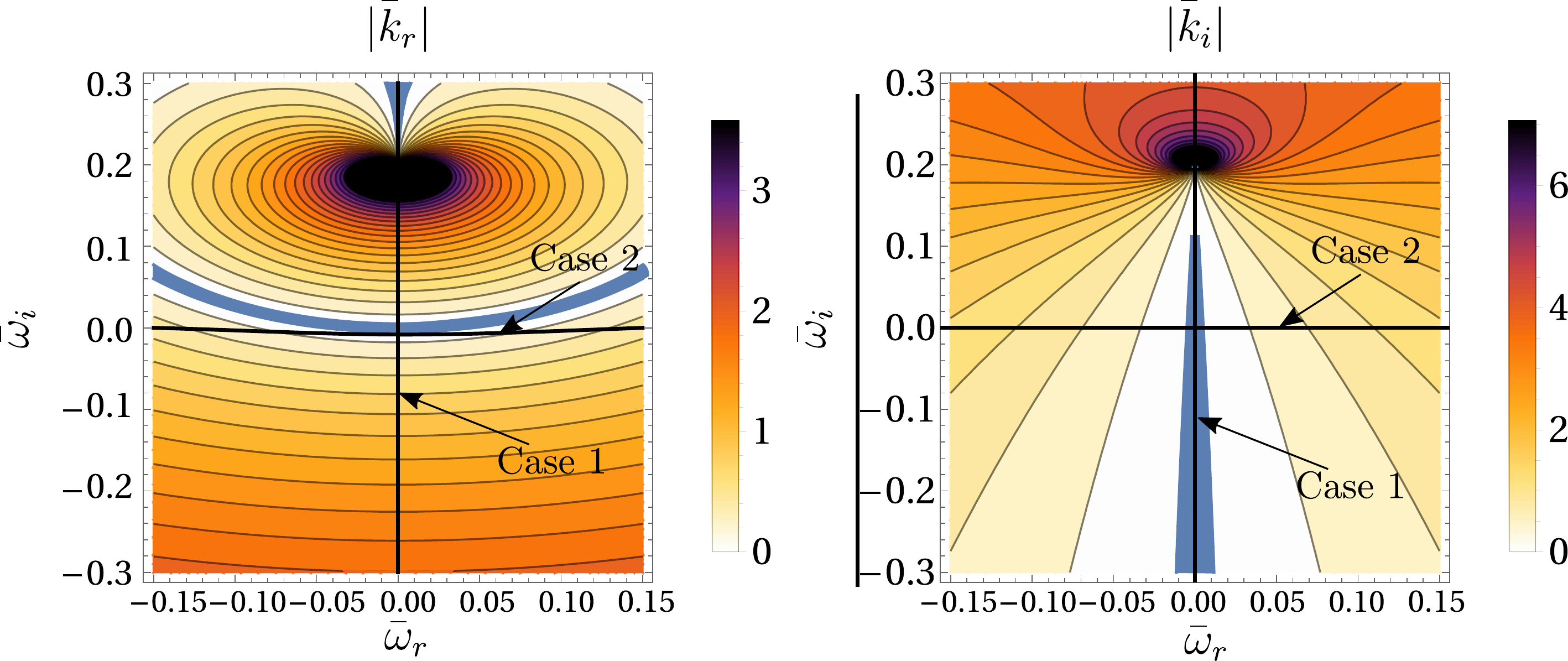}
    \caption{Contourplot of $\bar{k}_r,\;|\bar{k}_i|$ for values of $\bar{\omega}_r,\;\bar{\omega}_i$ close to the pole value $\bar{\omega}^P=-0.2i$. The contour closer to $\bar{k} \rightarrow 0$ is presented with blue color.}
    \label{fig: dispersion_kr_ki}
\end{figure}


\subsubsection{Case 1: Standing waves \label{Case 1}}
\noindent We focus our attention on the line where $\text{Re}[\bar{\omega}]=\bar{\omega}_r=0$ (see also Figure \ref{fig: dispersion_kr_ki}, Case 1). 
In this case standing waves are present in the medium. The amplitude of these standing waves is dependent on the values of $\bar{k}_i, \bar{\omega}_i$. When $\bar{k}_i>0$, while $\bar{\omega}_i=0$, the amplitude of the standing wave decreases with the distance from the origin as shown in equation \eqref{ubar_expanded}. However, if $\bar{\omega}_i>0$, then the value of the amplitude of the oscillations at fixed positions, will grow exponentially with time. Thus strain localization will happen inside a length $\bar{\lambda}=\frac{2\pi}{\bar{k}_r}$.\\ 

\begin{figure}[h]
    \centering
    \includegraphics[width=0.9\textwidth]{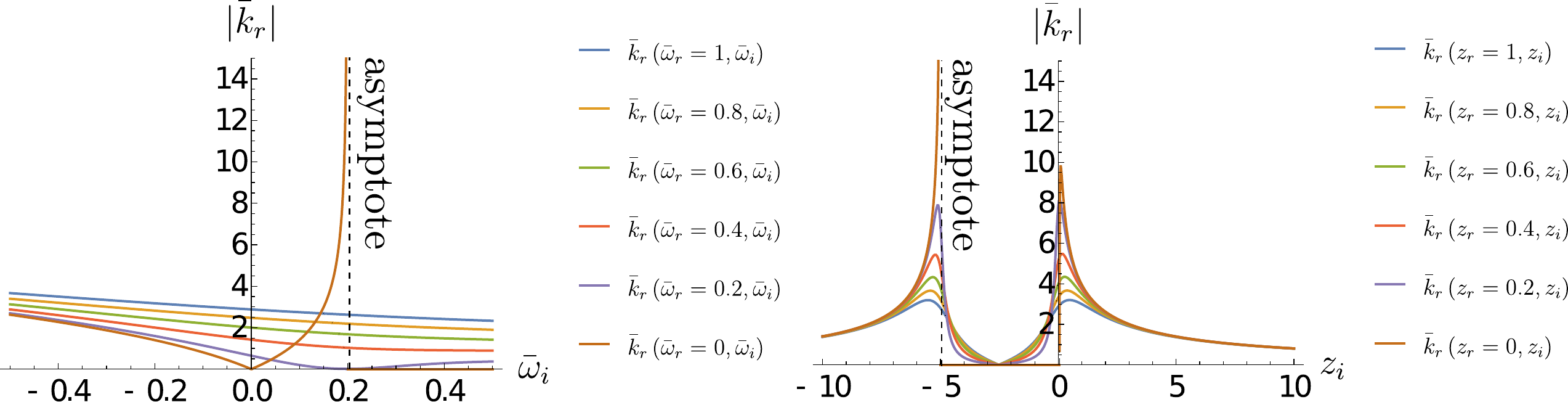}
    \caption{Left: Evolution of $\bar{k}_r$ with respect to $\bar{\omega}_i$ for parameter values $\bar{\omega}_r$  close to the pole value $\bar{\omega}^\text{P1}=0.2i$ indicating traveling waves around the pole $(\bar{\omega}_r\neq 0)$. Right:  $|k_r|$ along the lines of $\bar{\omega}_r=const$ for a range of values of $\bar{\omega}_i$ close to infinity.For $\bar{z}_r=0$ the imaginary axis $\bar{\omega}_i$ is parallel to the imaginary axis $z_i$. Therefore The detail around $z_i=0$ is indicative of the behavior of function $\bar{k}({\bar{\omega}})$ as $\bar{\omega}_r=0,\;\bar{\omega}_i\rightarrow \infty$.}
    \label{fig: dispersion_kr_omegai}
\end{figure}

\begin{figure}[h]
    \centering
    \includegraphics[width=0.9\textwidth]{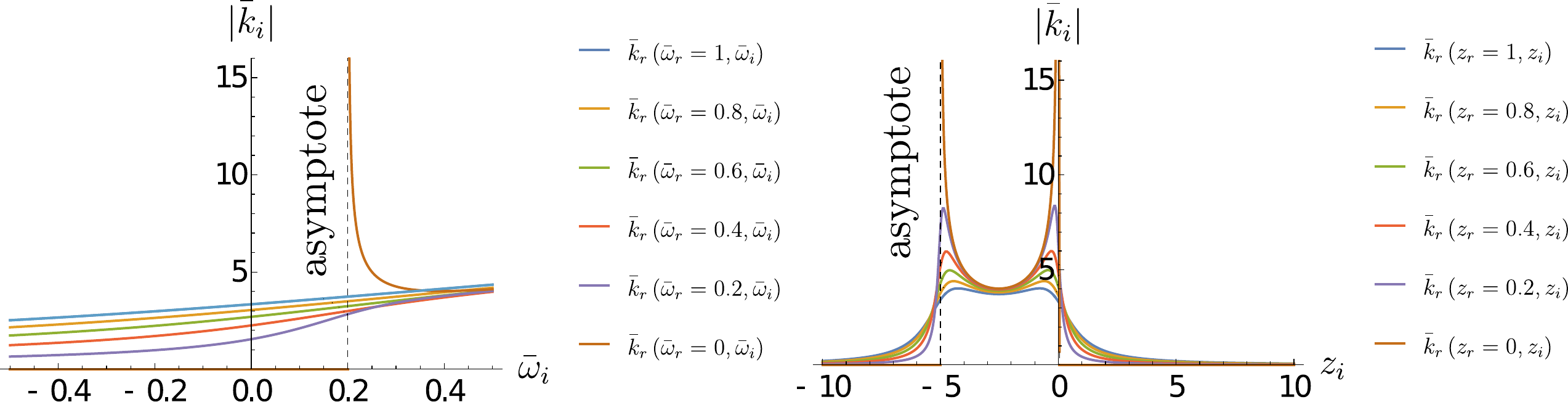}
    \caption{Evolution of $\bar{k}_i$ with respect to $\bar{\omega}_i$ for parameter values $\bar{\omega}_r$  close to the pole value $\bar{\omega}^\text{P1}=0.2i$ indicating traveling waves around the pole $(\bar{\omega}_r\neq 0)$. Left: $|\bar{k}_i(\bar{\omega})|$ along the lines of $\bar{\omega}_r=const$ for a range of values of $\bar{\omega}_i$, detail around the pole region $\bar{\omega}^\text{P1}$. Right: $|\bar{k}_i(\frac{1}{z})|$ along lines of constant $z_r$. For $z_r=0$ the imaginary axis $\bar{\omega}_i$ is parallel to the imaginary axis $z_i$. Therefore The detail around $z_i=0$ is indicative of the behavior of function $\bar{k}({\bar{\omega}})$ as $\bar{\omega}_r=0,\;\bar{\omega}_i\rightarrow \infty$.}
    \label{fig: dispersion_ki_omegai}
\end{figure}

\noindent Along the imaginary axis ($\bar{\omega}_r=0$), there is only one position where $\bar{k}_r\rightarrow \infty$. This is the pole at $\bar{\omega}^{\text{P1}}$ which constitutes a branch point of the dispersion relation \eqref{dispersion_relation_main} (see also Figure \ref{fig: dispersion_kr_omegai}-asymptote). In this point $\bar{k}_r(\bar{\omega}^{\text{P1}})\rightarrow \pm\infty$ as well as $\bar{k}_i(\bar{\omega}^{\text{P1}})\rightarrow \pm\infty$. We notice that the value of $\bar{\omega}_i$ at the pole is a cutoff value. There are no higher values of $\bar{\omega}_i$ for which standing waves are possible since $\bar{k}_r=0$ for $\bar{\omega}_i>\bar{\omega}^{\text{P1}}$(see also Figure \ref{fig: dispersion_kr_omegai}). For a perturbation from the initial homogeneous state to grow with time we are interested only in $\bar{\omega}_i>0$. Therefore the behavior of $\bar{k}_r$ for $\bar{\omega}_i<0$ is of no consequence for the stability of the homogeneous deformation, since all these modes will eventually die-off with time. From the above, we conclude that since the infinitesimal wavelength $\bar{\lambda}=\frac{2\pi}{\bar{k}_r}\rightarrow 0$ for the highest possible value of $\bar{\omega}_i$ strain localization on a mathematical plane is inevitable. \\
\newline
\noindent Next, we investigate the influence of $\bar{k}_i$ to the evolution of the amplitude of the perturbation. Looking at Figure \ref{fig: dispersion_ki_omegai} and considering the solution branch of equation \eqref{k_solution} we notice that $\bar{k}_i\rightarrow \infty$ for $\bar{\omega}=\bar{\omega}^{\text{P1}}$. Therefore, a standing wave with solution parameters $(\bar{\omega},\bar{k})$ defined the same as the pole $\bar{\omega}^{\text{P1}}$ will exhibit localization on a mathematical plane as distance $\bar{x}$ from the origin increases.

\subsubsection{Case 2: Traveling waves of zero temporal attenuation \label{Case 2}}
\begin{figure}[h]
    \centering
    \includegraphics[width=0.9\textwidth]{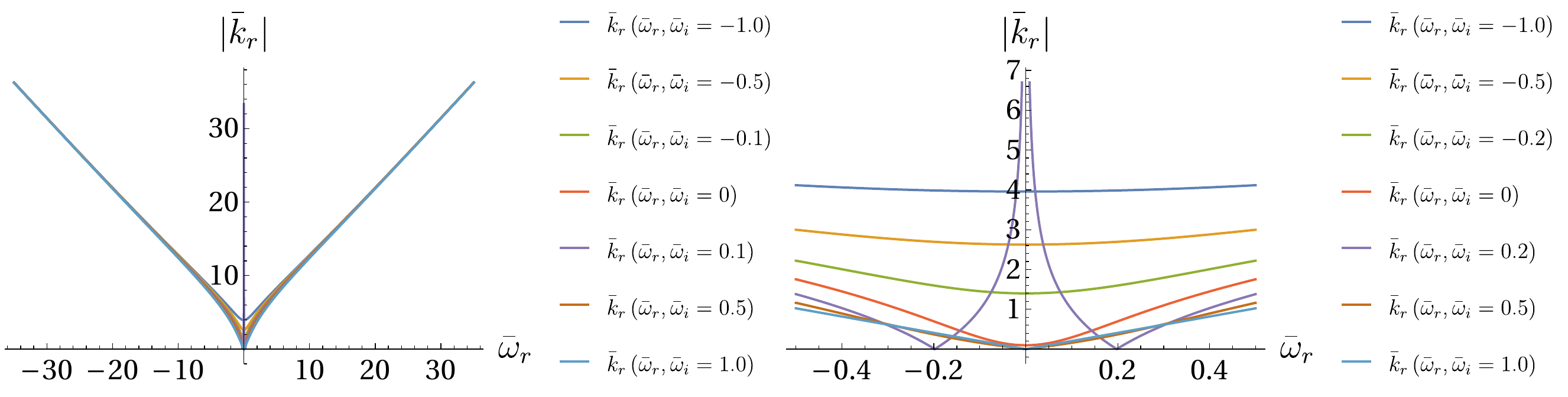}
    \caption{Dispersion curves $(\bar{\omega}_r,\bar{k}_r)$ for different values of parameter $\bar{\omega}_i$ along the line of zero temporal coefficient $\bar{\omega}_i=0$ (case 2). With red color and the value passing from the pole $\bar{\omega}^{\text{P1}}=0.2i$ purple color. Detail of the dispersion for low values of $\bar{\omega}_r$ is shown on left.}
    \label{fig: dispersion_kr_omegar}
\end{figure}

\noindent Another important case seen in bibliography \cite{Abellan2006,DeBorst2020,Sluys1992,wang1996interaction,wang1997viscoplasticity} is that of the traveling waves where the imaginary part of angular frequency is zero $\bar{\omega}_i=0$ (see Figure \ref{fig: dispersion_kr_ki}, Case 2). Therefore the Lyapunov exponent is also zero $(s=-i\bar{\omega}_i=0)$. In this case the amplitude growth is dependent only on $\bar{k}_i$: $
\bar{u}(\bar{x},\bar{t})=\exp(-\bar{k}_i\bar{x})\exp[(\bar{k}_r\bar{x}-\bar{\omega}_r\bar{t})]
$.
\noindent The parameter $\bar{k}_i$ corresponds to the parameter $\alpha$  in \cite{DeBorst2020,Sluys1992,wang1996interaction,wang1997viscoplasticity} and its inverse $l=\alpha^{-1}$ is thought to constitute a critical length that is supposed to regularize the problem, damping the waves of higher wavenumber $\bar{k}_r$ and therefore avoiding strain localization on a mathematical plane. Here we show that in fact depending on the solution branch of equation \eqref{k_solution} the contribution of $\bar{k}_i$ to the solution's behavior can instead be positive, indicating amplification of perturbations of higher wavenumber $\bar{k}_r$. 
\noindent We focus our attention on Figure \ref{fig: dispersion_kr_omegar}. We follow the red line corresponding to $\bar{\omega}_i=0$. We notice that the dispersion relation predicts $\bar{k}_r= 0$ for $(\bar{\omega}_r,\bar{\omega}_i)=(0,0)$. As we move away from the origin along the direction of $\bar{\omega_r}$ we notice a quasi-linear increase of the wavenumber $\bar{k}_r$. Figure \ref{fig: dispersion_kr_omegar} shows that $|\bar{k}_r|$ increases monotonically and tends to infinity for $|\bar{\omega}_r|\rightarrow \infty$. The latter can be proven mathematically (see Appendix \ref{Appendix_B}).
From this we establish that perturbations whose wavelength tends to zero $\bar{\lambda}\rightarrow 0$ are admissible. Next we proceed on examining the rate of increase of their amplitude with respect to distance $\bar{k}_{i}$.

\begin{figure}[!hbp]
    \centering
    \includegraphics[width=0.9\textwidth]{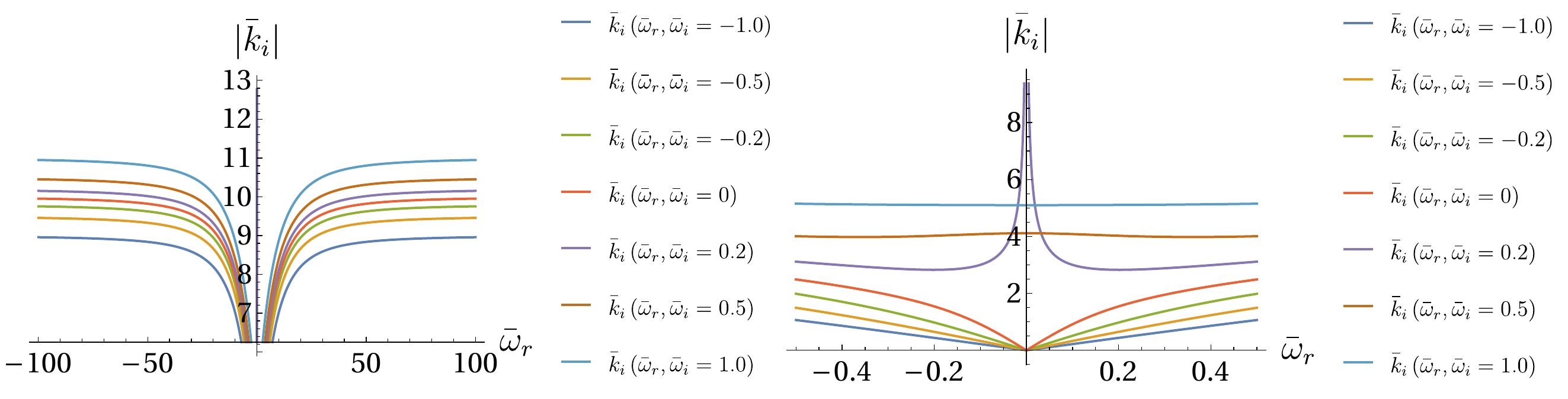}
    \caption{Evolution of $\bar{k}_i$ for different values of temporal coefficient $\bar{\omega}_i$ including the case zero temporal attenuation/amplification $\bar{\omega}_i=0$ to the pole value $\bar{\omega}^{\text{P1}}=0.2i$. For traveling waves around the pole $(\bar{\omega}_r\neq 0)$ the spatial attenuation coefficient $\bar{k}_i$ is reaching an upper bound. Subfigure on left presents the curve of $|k_i|$ along the lines of $\bar{\omega}_i=const$ for a range of values of $\bar{\omega}_r$ while subfigure on right presents a detail around the pole region $\bar{\omega}^\text{P1}$.}
    \label{fig: dispersion_ki_omegar}
\end{figure}
  
\noindent 
\noindent Figure \ref{fig: dispersion_ki_omegar} shows that for $\bar{\omega}_r$ tending to infinity, the value of $|\bar{k}_i|$ increases monotonically and tends to a ceiling value $\bar{k}_i\rightarrow c\;\in \mathbb{R}$. The latter can be proven mathematically (see Appendix \ref{Appendix_B}). Therefore, since when $\bar{k}_r(\omega)\rightarrow \infty$, $\bar{k}_i<0$ takes its maximum absolute value (see Figures \ref{fig: dispersion_kr_omegar}, \ref{fig: dispersion_ki_omegar}), we notice that the amplitude of the perturbation of zero wavelength $\bar{\lambda}$ is increasing the fastest as the perturbation travels through the medium. Therefore, strain localization on a \textit{traveling mathematical plane} will happen. 

\subsubsection{Case 3: The general case of traveling waves,\label{Case 3}}
\noindent Based on Figure \ref{fig: dispersion_kr_ki} and the diagrams of Figures   
\ref{fig: dispersion_kr_omegai},  
\ref{fig: dispersion_ki_omegai}, 
\ref{fig: dispersion_kr_omegar} and
\ref{fig: dispersion_ki_omegar} very general cases of traveling waves can be examined. In Figures \ref{fig: dispersion_kr_omegai} and  
\ref{fig: dispersion_ki_omegai} we examine the evolution of $\bar{k}_r$, $\bar{k}_i$ with respect to the imaginary part of the frequency $\bar{\omega}_i$, by considering the real part of the angular frequency $\bar{\omega}_r$ as a parameter.\\
\\
\noindent We already presented case 1 of standing waves $\bar{\omega}_r=0$ where the influence of the pole leads to strain localization  due to $\bar{\omega}_i>0$ for $\bar{k}_r(\bar{\omega})\rightarrow \infty$ as seen on the left diagram of  Figure \ref{fig: dispersion_kr_omegai}. For the rest of the values of the parameter $\bar{\omega}_r$, the wavenumber $\bar{k}_r$ is bounded. Therefore, no strain localization on a mathematical plane will take place in these cases.\\
\newline
\noindent In Figures \ref{fig: dispersion_kr_omegar} and \ref{fig: dispersion_ki_omegar} temporal amplification $\bar{\omega}_i$ is introduced as a parameter, keeping $\bar{\omega}_r$ as the independent variable. Figure \ref{fig: dispersion_kr_omegar} is indicative of the dispersion relation of the medium. 
\noindent Away from $\bar{\omega}_r=0$ the dispersion relation $\bar{\omega}_r,\bar{k}_r$ becomes linear for all values of $\bar{\omega}_i$ and the resulting traveling waves have a common phase velocity. We notice here that as $\bar{\omega}_r \rightarrow \infty$, $\bar{k}_r(\bar{\omega}_r,\bar{\omega}_i) \rightarrow \infty$. For the waves with the same value for the parameter $\bar{\omega}_i>0$ this means that the growth  of their amplitude in time is the same. 
\noindent 
For large real angular frequencies $\bar{\omega}_r$ the spatial amplification coefficient $|\bar{k}_i|$ presents a ceiling value. This result is already proven for $\bar{\omega}_i=0$ (see Appendix \ref{Appendix_B}). The ceiling value depends on the parameter value $\omega_i$, namely it increases as the parameter $\bar{\omega}_i$ increases. The ceiling value of the amplification coefficient $\bar{k}_i$ corresponds to a wavenumber $\bar{k}_r$ that tends to infinity $\bar{k}_r\rightarrow \infty$. This result shows that in the general case of traveling waves with infinitesimal wavelength $\bar{\lambda}=\frac{2\pi}{\bar{k}_r}\rightarrow 0$, strain localization on a traveling mathematical plane will happen due to the combination of $\bar{k}_i<0$ and $\bar{\omega}_i>0$, provided that $\bar{k}_r \rightarrow \infty$.\\
\newline
\noindent In the above we focused mainly on limit cases related to strain localization in a elasto-viscoplastic strain-softening $(h<0)$, strain rate-hardening medium $(g>0)$. For a more general, qualitative description of a traveling monochromatic pulse, we refer to Appendix \ref{sec: appendix C}. In this Appendix we refer also to the connection between strain localization and the interplay between phase and amplitude velocities.
  
\subsubsection{Behavior of the pole at infinity $\omega^{\text{P2}}$ \label{Behavior_P2}}

\noindent Based on the behavior close to infinity, $\bar{\omega}\rightarrow\infty$ or $z\rightarrow 0$, we get $\bar{k}(\bar{\omega})=\bar{k}(\frac{1}{z})\rightarrow\infty$. In contrast to real $\infty$ that can be either positive or negative or indeterminate based on whether we approach the value of $z$ from above or below zero, complex $\infty$ is indeterminate as at the pole value $\bar{\omega}^{P2}$ the limit along each direction surrounding the pole indicates differences in the real and imaginary parts of $\bar{k}(\bar{\omega})$. Since around a simple pole like $\bar{\omega}^{\text{P2}}$, where $\bar{\omega}^\text{P2}_r\rightarrow\infty,\;\bar{\omega}^\text{P2}_i\rightarrow\infty$ the argument of a complex function changes by a full $2\pi$ radians, the two limiting cases for the value of $\bar{k}(\bar{\omega}^{\text{P2}})$ are $\text{Re}[\bar{k}]=\bar{k}_r\rightarrow \infty,\text{Im}[\bar{k}]=\bar{k}_r\rightarrow 0$ and $\text{Re}[\bar{k}]=\bar{k}_r\rightarrow 0,\text{Im}[\bar{k}]=\bar{k}_r\rightarrow \infty$.  Since $\bar{k}_r \rightarrow \infty$ when $\bar{\omega}=\bar{\omega}^{\text{P2}}$, we conclude again, that localization on a mathematical plane is possible. Since, localization on a mathematical plane happens for values of $\bar{\omega} \rightarrow \infty$, the rate of increase of the perturbation amplitude as given by the Lyapunov exponent $\bar{s}=-i\bar{\omega}_i$ is unbounded.\\

\subsubsection{Influence of the pole $\bar{\omega}^{P1}$ \label{Behavior_P1}}
\noindent By expanding the solution space allowing for complex $\bar{\omega}_i$ we allow a connection with the Lyapunov exponent $\bar{s}$ used in stability analyses. The new solution space is richer regarding the perturbations we can introduce in the visco-elastoplastic medium. Some key characteristics retained by the solution  from the definitions already found in the literature, is the exclusion of standing waves of infinitesimal length that grow with an infinite Lyapunov coefficient as in the case of the pure ill-posed rate-independent plasticity problem.\\
\newline
\noindent The introduction of the parameter $T=\frac{\bar{\eta}^{vp}}{t_c}$ allows for the existence of a zero value on the imaginary axis. This zero in turn plays the role of the branch point, forcing the argument of the pole $\bar{\omega}^{P2}$ at infinity to turn by $\pi/2$ thus nullifying the real part of $\bar{k}(\bar{\omega})$ along the imaginary axis. If that zero was not there, then the point at infinity would be a branch point, therefore strain localization on a mathematical plane would happen for infinite $\bar{\omega}_i$ as in the case of a strain-softening rate-independent material.\\
\newline
\noindent The visco-elastoplastic medium discussed here under the expansion of its dispersion equation solution negates the instantaneous localization of deformation, however as dictated by the pole $\bar{\omega}^{P1}$ amplification of the infinitesimal wavelength perturbation is still possible. In other words, the value of the Lyapunov exponent $\bar{s}=-i\bar{\omega}_i$ becomes bounded due to viscoplasticity but this is not true for the value of the wavenumber $\bar{k}_r$. 
\subsubsection{Comparison to case M1 (stable configuration) \label{sec: Theory stable configuration}}

\noindent For case M1 we note that $\bar{h}=0.03>0$. In this case the points of interest of the dispersion relation \eqref{final_form_k1} change leading to different behavior than the one previously presented. The relationships for the determination of zeros and poles of the function remain the same (see section \ref{Poles_and_zeros}).


\noindent For these numerical values, the pole $\bar{\omega}^{\text{P1}}$ is reflected due to the change of sign of $\bar{h}$ around the Real axis. This however is not true for the he zero $\bar{\omega}^{\text{O1}}$ since the sign of $(-3+h)$ does not change (provided that $|\bar{h}|<1$).  This change will move the branch cut defined on the imaginary axis below the origin $\bar{\omega}^{\text{O1}}$.  Since now the pole lies on $\bar{\omega}_i<0$ the perturbations of infinitesimal length corresponding to it $\bar{\lambda} \rightarrow 0,\bar{k}_r\rightarrow \infty$ are attenuated with time, therefore strain localization on a mathematical plane is not possible in this case.

\section{Numerical analysis \label{sec: Numerical_Analysis}}

\noindent In this section, two sets of numerical analyses are performed in order to verify the stability and possible strain localization of the elasto-viscoplastic strain softening $(h<0)$, strain-rate hardening $(g>)0$ wave equation. Two sets of analyses were performed. In the first set, a 1D model of an elasto-viscoplastic infinite string was used on which, the linear differential equation of third order (equation \eqref{normalized equation_main}) is numerically solved for a variety of initial conditions. We comment on the displacement profiles that develop with time and we verify the theoretical findings of section \ref{sec: three}. In the second set, we proceed in numerically solving the fully non-linear problem. The difference between these two sets of analyses, lies in the fact that in the second case unloading is allowed to take place. Therefore, we can investigate its influence on the strain localization profiles. The non linear numerical analyses show also strain localization and mesh dependency.

\subsection{Linearized model: Model description \label{sec: Linearized model: Model description }}
\noindent To model the infinite string a large length and the Sommerfield open boundary conditions were used. The latter can be used since the partial differential equation in question is linear and by use of the Fourier transform can be shown to have partial solutions in the form of $A\exp (i(\bar{\omega} \bar{t}-\bar{k}\bar{x}))$. Three modes of inducing the perturbation from the reference homogeneous state were examined making use of non zero initial conditions for the string displacement. This is achieved by varying the shape of the perturbation with three different ways: an initial pinch of the string at the middle, a cosinus pulse centered in the middle as well as a Gaussian distribution centered at the middle.\\
\newline 
\noindent Two sets of parameters were used for the analyses, where the sign of the hardening parameter $\bar{h}$ varies between positive or negative in order to compare between strain-hardening and a strain-softening material as shown in Table \ref{Table: material_params}. The material parameters used in the unstable case are those provided by \cite{DeBorst2020}. The numerical analyses were performed using the method of Finite Differences. In particular a central difference scheme was used for the spatial discretization of the PDE problem, the domain of length $L=15$m is discretized into 250 segments, resulting in a coupled system of ODE's which was solved by the algorithm IDA of the Mathematica$^{TM}$ software package \cite{Mathematica}.   

\subsubsection{Pinching}
\begin{figure}[h]
    \centering
    \begin{minipage}{.5\textwidth}
    \centering
    \includegraphics[width=0.9\linewidth]{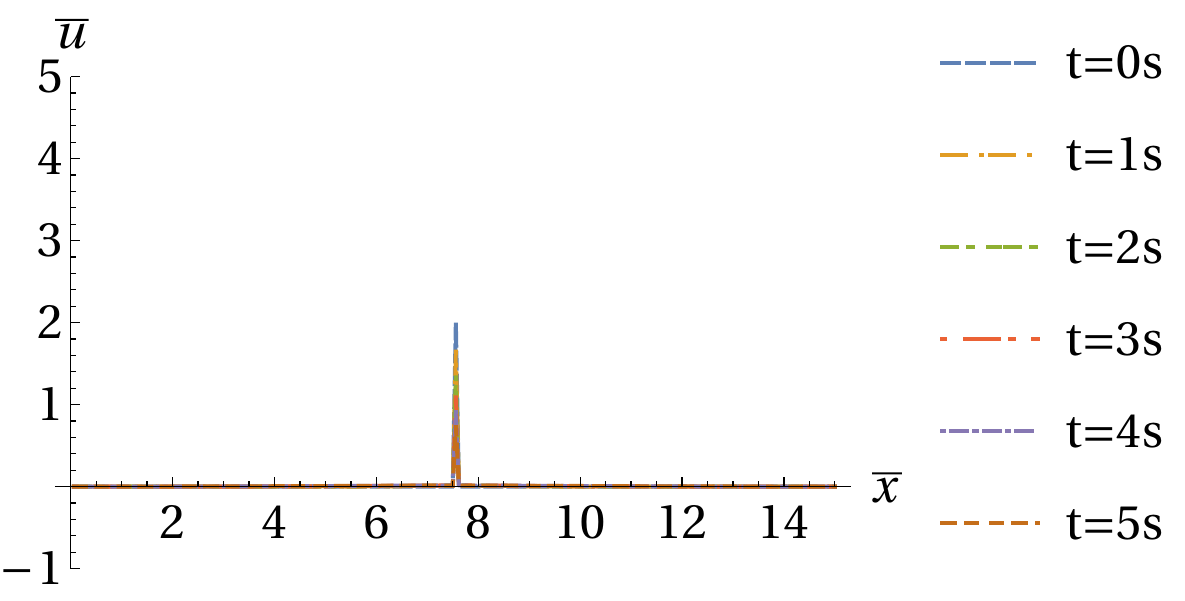}
    \end{minipage}%
    \begin{minipage}{.5\textwidth}
    \centering
    \includegraphics[width=0.9\linewidth]{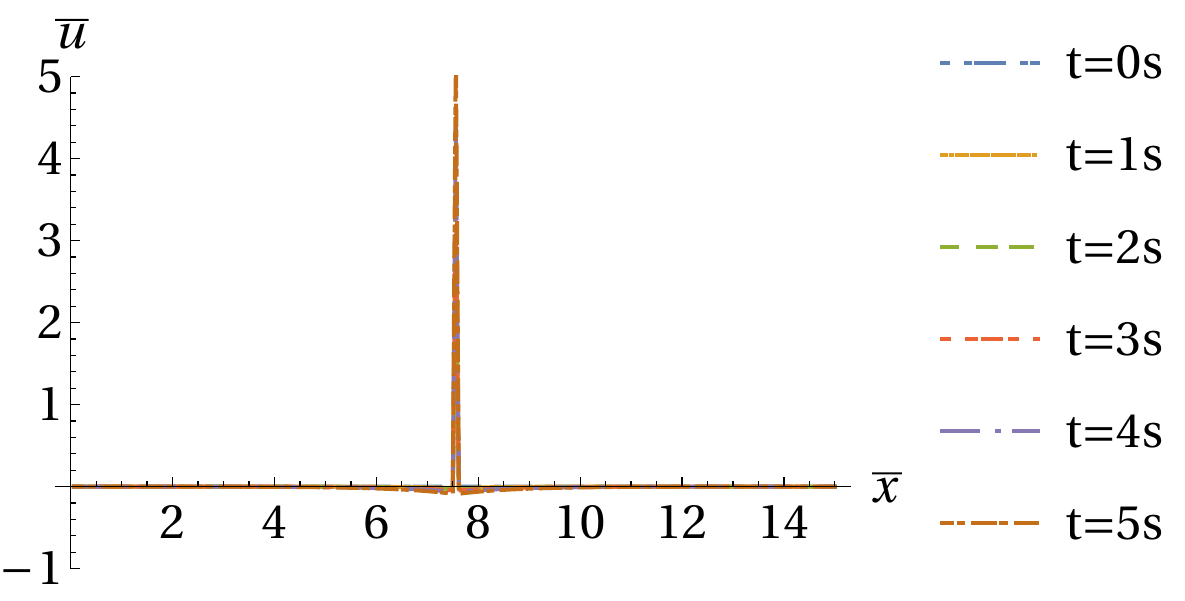}
    \end{minipage}\qquad
    \caption{Evolution of the pinching perturbation at different times. Left: strain hardening material case M1. On the right: strain softening material case M2.}
    \label{fig: pinch_stable_unstable}
\end{figure}

\noindent We present in Figure \ref{fig: pinch_stable_unstable} the behavior of the string after an initial pinching -application of initial displacement conditions at the middle node of the discretized domain. Mesh convergence analysis has shown that, in the unstable case M2, the localization width is equal to the mesh size. 
Therefore, the elasto-viscoplastic formulation   
does not regularize the underlying problem as presented in the introduction and the solution is mesh dependent. This is in accordance with the theoretical results presented in subsections \ref{Case 1}, \ref{Case 2}, \ref{Case 3}. We note also that the strain hardening material of the M1 case does not lead to strain localization as expected (see section \ref{sec: Theory stable configuration}).

\subsubsection{Monochromatic cosine pulses}
\begin{figure}[h]
    \centering
    \begin{minipage}{.45\textwidth}
    \centering
    \includegraphics[width=0.9\linewidth]{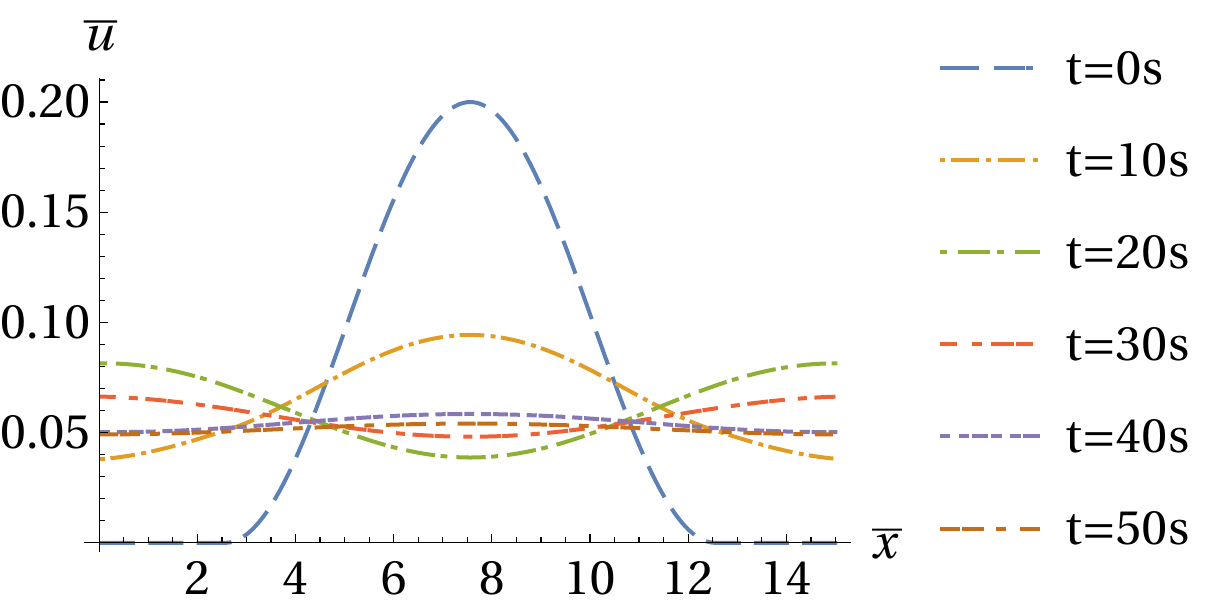}
    \end{minipage}\qquad
    \begin{minipage}{.45\textwidth}
    \centering
    \includegraphics[width=0.9\linewidth]{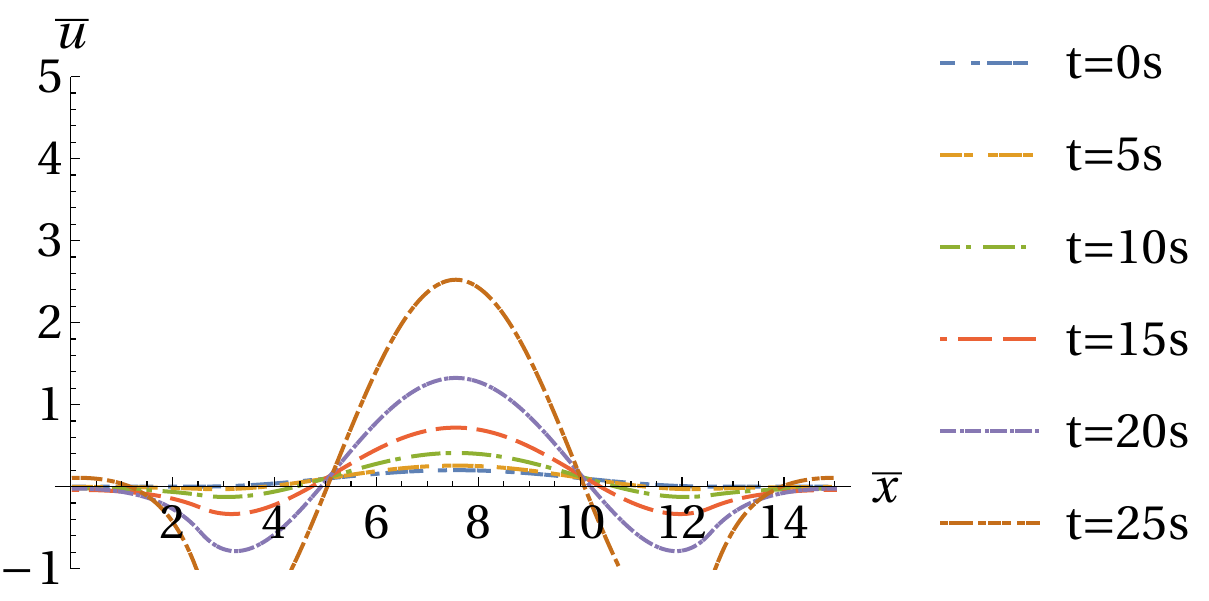}
    \end{minipage}%
    \caption{Evolution of the cosine perturbation at different times. Left: strain hardening material case M1. On the right: strain softening material case M2.}
    \label{fig: cosine_stable_unstable}
\end{figure}

\noindent The behavior of the string after application of cosine initial conditions for the two sets of parameters (see Table \ref{Table: material_params}) is presented in Figure \ref{fig: cosine_stable_unstable}. Again we notice a localization of deformation for the case of strain softening.\\

\begin{figure}[h]
    \centering
    \includegraphics[width=0.7\linewidth]{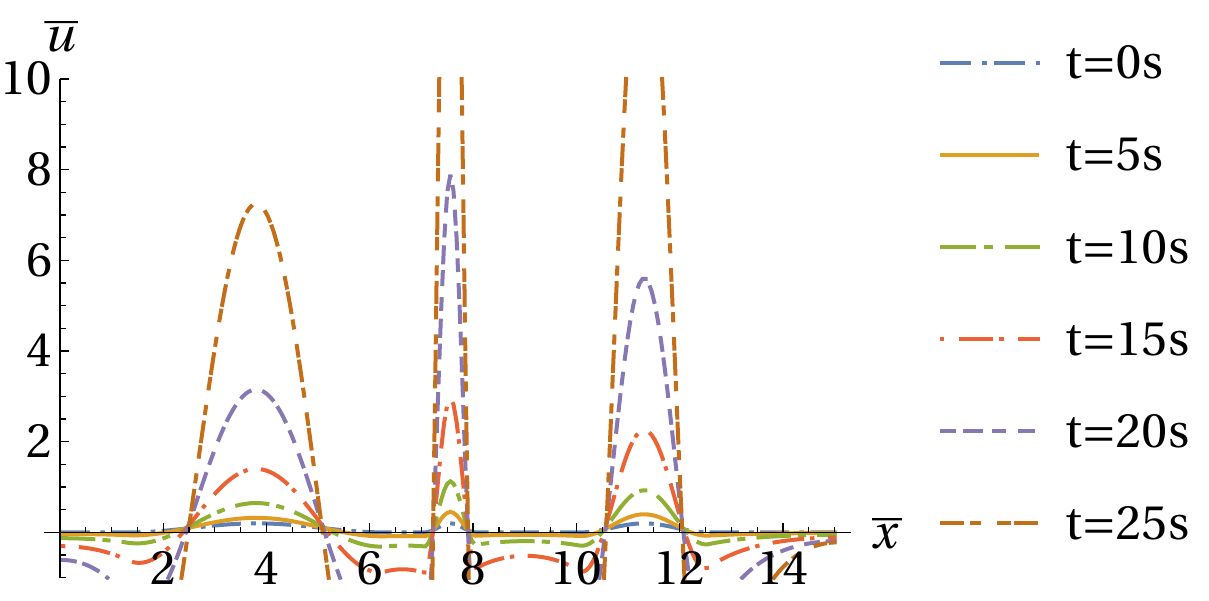}
    \caption{Evolution of the different wavelength cosine perturbations. Faster strain localization is observed for the smallest wavelength.}
    \label{fig: cos_pert_compare}
\end{figure}

\noindent In order to verify the theoretical prediction, that the elasto-viscoplastic medium with strain softening localizes on a mathematical plane under dynamic loading conditions, we superpose three different cosine perturbations in the medium by varying the width of each perturbation as shown in Figure \ref{fig: cos_pert_compare}. The perturbation wavelengths are $\bar{\lambda}$: 1, 5, 10, corresponding to perturbation widths of 0.5 ,2.5 , 5. In Figure \ref{fig: cos_pert_compare}, the position of the perturbations from left to right is $\bar{x}=3.75 \rightarrow \bar{\lambda}=10,\; \bar{x}=7.5 \rightarrow \bar{\lambda}=1 \;\text{and}\;\bar{x}=11.24 \rightarrow \bar{\lambda}=5$ respectively. Figure \ref{fig: cos_pert_compare} shows that localization is accumulating faster for the smallest perturbation length, verifying the theoretical findings of section $\ref{sec: three}$.    

\subsubsection{Centered Gaussian initial condition}
\begin{figure}[h]
    \centering
    \begin{minipage}{.45\textwidth}
    \centering
    \includegraphics[width=0.9\linewidth]{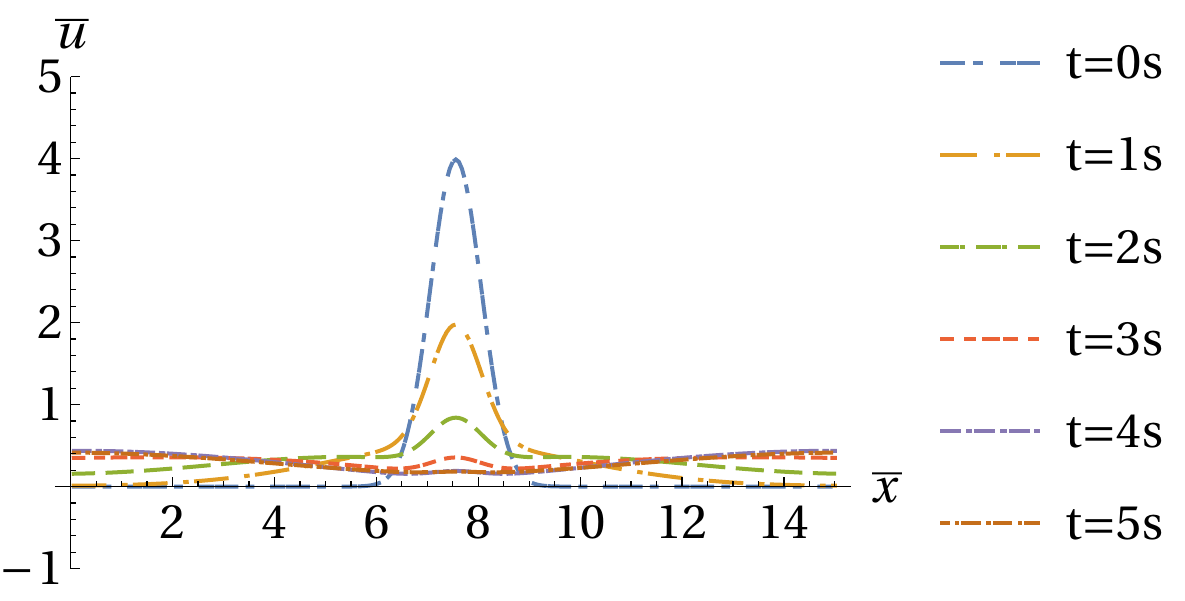}
    \end{minipage}\qquad
    \begin{minipage}{.45\textwidth}
    \centering
    \includegraphics[width=0.9\linewidth]{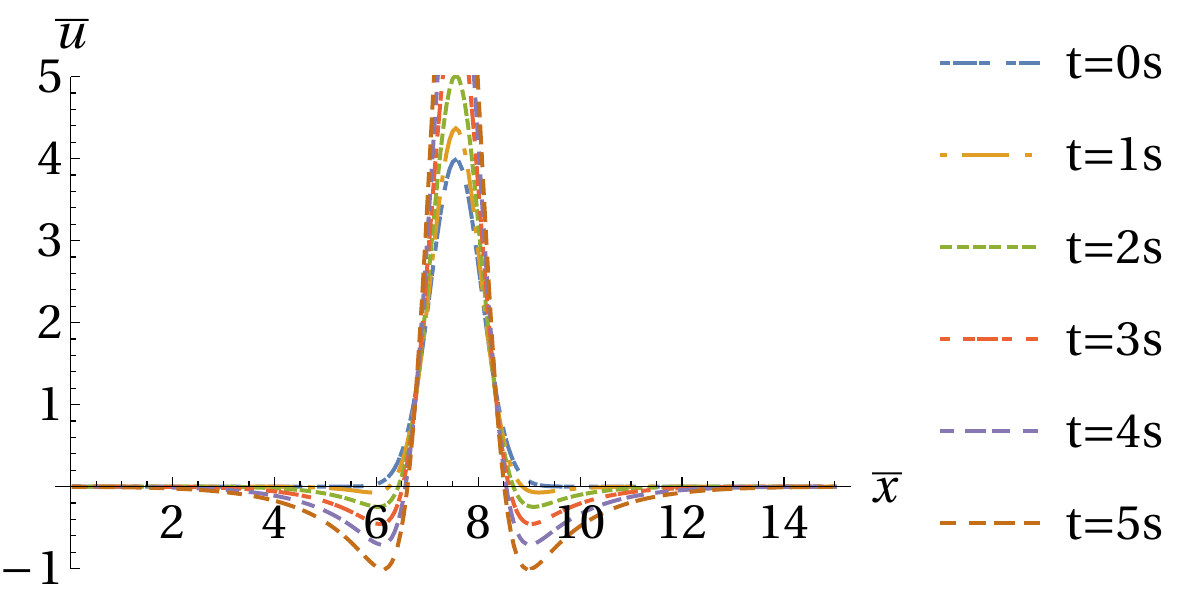}
    \end{minipage}%
    \caption{Evolution of the Gaussian perturbation at different times. Left: strain hardening material case M1. On the right: strain softening material case M2.}
    \label{fig: gaussian_stable_unstable}
\end{figure}

\begin{figure}[h]
    \centering
    \includegraphics[width=0.7\linewidth]{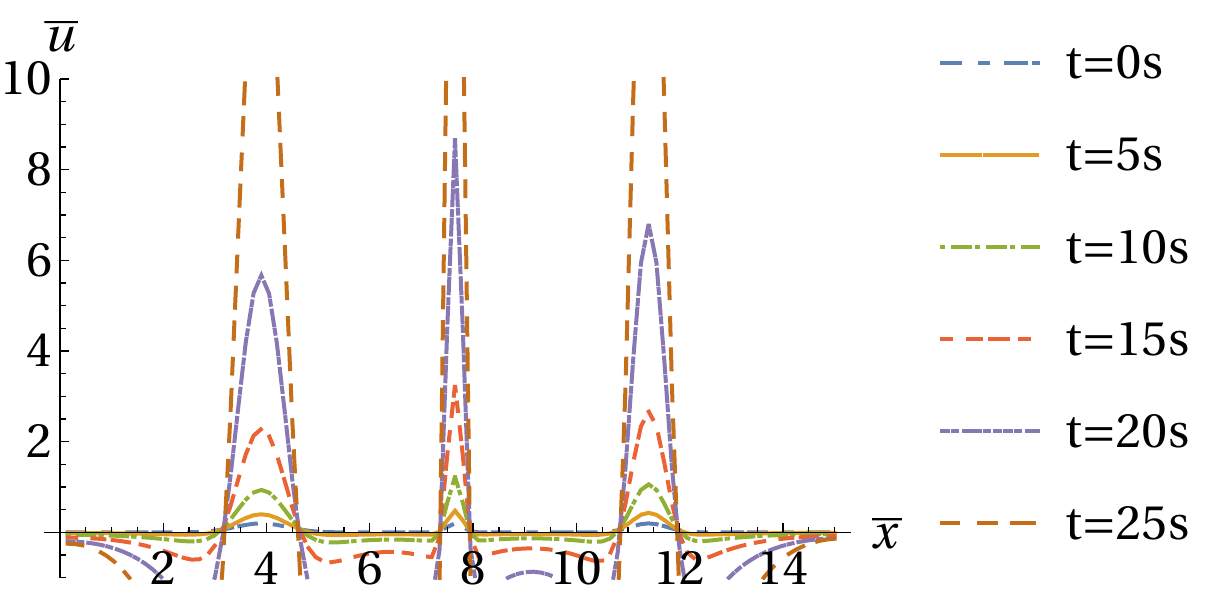}
    \caption{Evolution of different Gaussian perturbations varying in their width. Faster strain localization is observed for the smallest wavelength.}
    \label{fig: gauss_pert_compare}
\end{figure}

\noindent In Figure \ref{fig: gaussian_stable_unstable}  we present the behavior of the string after a gaussian perturbation of the initial conditions -application of initial gaussian displacement conditions centered at the middle node of the discretized domain. In the strain softening case M2, we notice that the localization of the deformation is contained into a narrow band of finite length, dependent on the width of the initial perturbation as shown on the right of Figure \ref{fig: gaussian_stable_unstable}.\\ 
\newline
\noindent We emphasize, that the lower bound of localization is a result of the mesh discretization. Further increase of the mesh will lead to narrower bands as expected by theory mesh dependency. In Figure \ref{fig: gauss_pert_compare} we compare among the perturbation profiles at different times for Gaussian perturbations of varying width. As in the previous case, the narrowest perturbation localizes the fastest.  

\subsection{Fully non-linear problem}

\begin{figure}[H]
    \centering
    \includegraphics[width=0.9\linewidth]{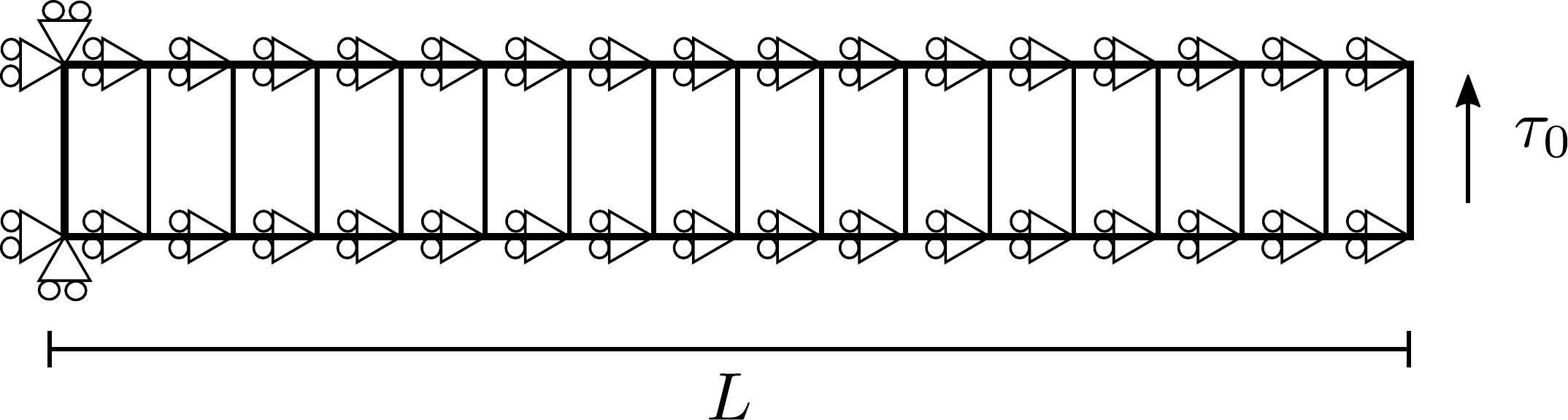}
    \caption{2D model of a layer subjected to shear.}
    \label{fig: shear_layer}
\end{figure}

\noindent In order to model the effect of viscous regularization on the strain localization and possible mesh dependence  in fully nonlinear cases without neglecting the effect of unloading, a series of dynamic numerical analyses was performed using the finite element analysis program ABAQUS (\cite{0b112d0e5eba4b7f9768cfe1d818872e}). The fully implicit Newmark scheme was used. 
The parameters used for the Newmark scheme correspond to the trapezoidal rule ($\alpha=0,\;\beta=\frac{1}{4},\;\gamma=\frac{1}{2}$) in order to avoid numerical damping.\\
\newline
\noindent The use of a User Material Subroutine (UMAT) was favored in order to incorporate the Perzyna elasto-viscoplastic constitutive material law into ABAQUS. 
The material parameters leading to mesh independent solution were taken from \cite{DeBorst2020}, see Table \ref{Table: material_params2}. These values are quite low for real physical applications, but they are used following \cite{DeBorst2020} in order to allow direct comparisons. Configuration D1 corresponds to a set of material parameters that seems to lead to regularization of strain localization and therefore to mesh independent results. However, as it will be shown, this is not always the case. Configuration D2 corresponds to a set of parameters leading to strain localization and ,therefore, to mesh dependent results. Care was taken to remove additional viscosity from the analysis except for the one strictly prescribed by the material. The analyses were performed using 2D solid, reduced integration elements CPE4R.\\
\newline
\noindent We study the pure shear of a 1D layer of length $L$ (see Figure \ref{fig: shear_layer}). To avoid bending we block displacements along the length of the model. A shear traction $\tau_0=14$ Pa is applied instantaneously on top of the model and it propagates towards the fixed end at the base of the layer. We extract our results after the pulse has returned to the free end of the model. For the duration of the analysis the time increment is kept smaller than  $\Delta t = 0.001$ s, which is smaller than the time $\Delta_{CFL}$ needed for the elastic wave to traverse the smallest element dimension of the mesh as specified by the Courant-Friedrichs-Lewy (CFL) criterion. Considering the length of the model $L=1$ m, and the material parameters of Table \ref{Table: material_params2} leading to the elastic shear wave velocity $c_g=\sqrt{\frac{G}{\rho}}=4$ $\frac{m}{s}$, for a mesh discretization of 200 elements, we obtain:  $\Delta t_{\text{CFL}}=\frac{1}{200*4}=0.00125$ s. The time increment selected for all analyses is given in Table \ref{Table: time_increment_values}.
\\

\begin{table}
\begin{center}
\begin{tabular}{ l l l l } 
\hline
  analyses  & D1   & D2 &\\
  mesh independence &  Yes & No  & Units\\
 \hline
 $\rho$ & 1250 & 1250  &$\frac{kg}{m^3}$\\
 $G$ & 20000 & 20000 & Pa \\ 
 $H$ & -200 & -1000 & Pa \\ 
 $c$ & 20 & 20 & Pa\\ 
 $\eta^{vp}=\frac{g}{F_0}$ & 50 & 25 & s\\ 
 \hline
\end{tabular}
\caption{\label{Table: material_params2}Material parameters used for the ABAQUS numerical analyses. The first set of parameters is taken from \cite{DeBorst2020}.}
\end{center}
\end{table}

\begin{table}
\begin{center}
\begin{tabular}{ l l l l } 
\hline
  analyses (D1, D2)  & $\Delta t$  \\
  element number & [s] \\
 \hline
 25 & 0.001 &  \\
 50 & 0.001 &\\
 100 & 0.001 & \\ 
 200 & 0.001 \\ 
 400 & 0.0002\\ 
 800 & 0.0002\\ 
 \hline
\end{tabular}
\caption{\label{Table: time_increment_values}Time increment $\Delta t$ measured in seconds, used for the ABAQUS numerical analyses, such that the CFL yield criterion is satisfied for each mesh discretization.}
\end{center}
\end{table}

\begin{figure}[h]
    \centering
    \begin{minipage}{.37\textwidth}
    \centering
    \includegraphics[width=0.9\linewidth]{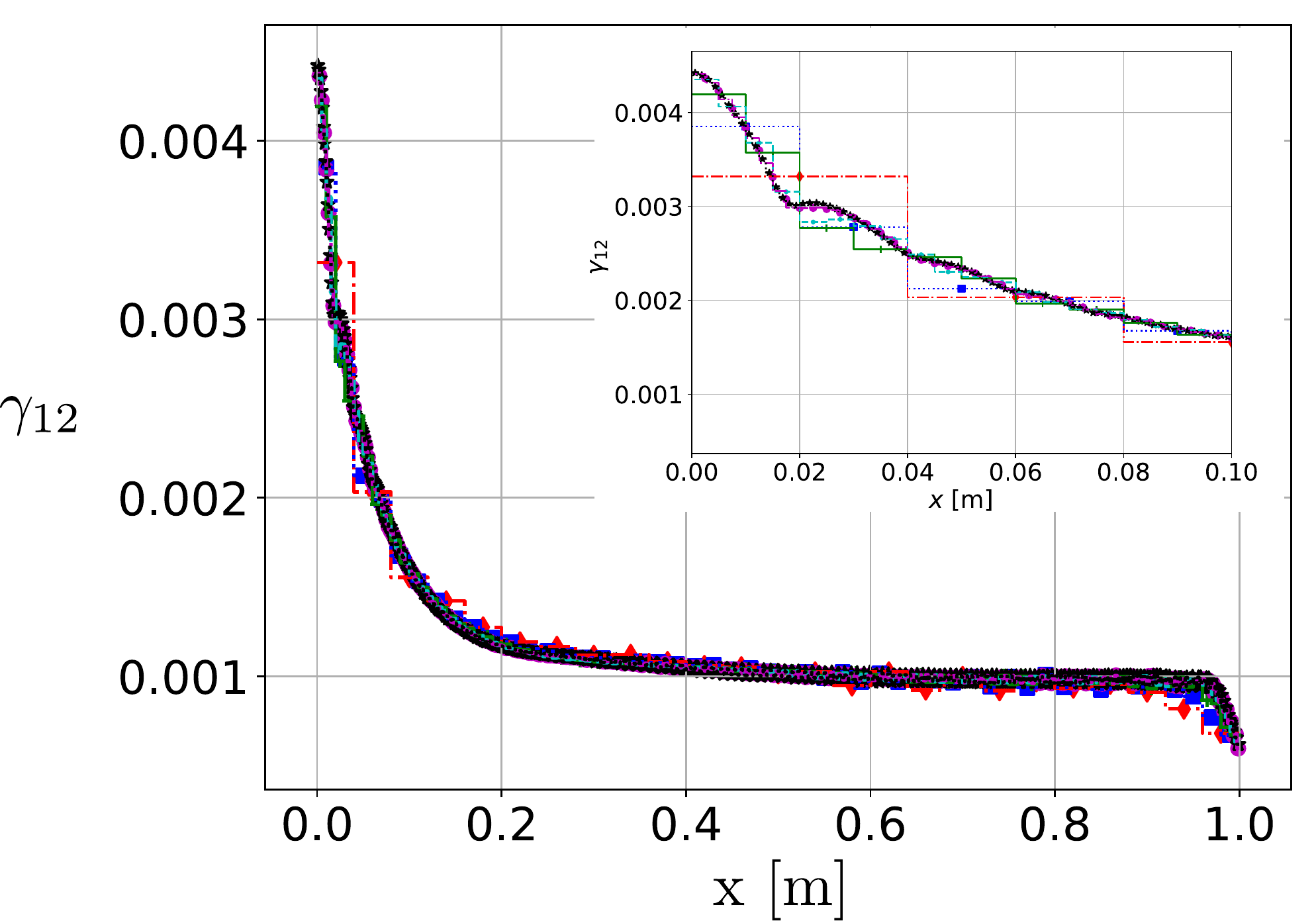}
    \end{minipage}\qquad
    \begin{minipage}{.53\textwidth}
    \centering
    \includegraphics[width=0.9\linewidth]{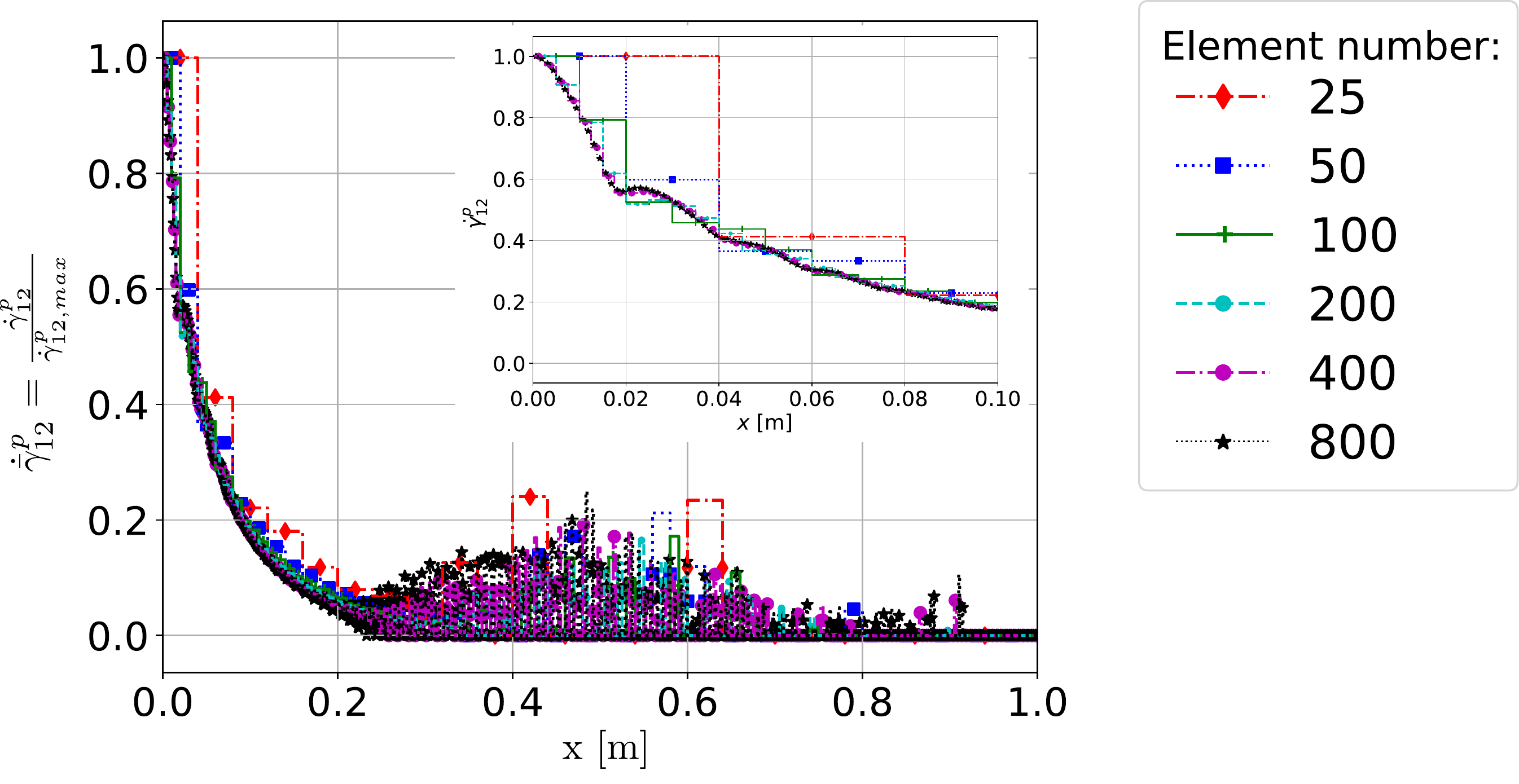}
    \end{minipage}%
    \caption{ On the left: Total shear strain $\gamma_{12}$ profiles at the end of the analysis for different mesh discretization of the model for the material set D1. On the right: Normalized plastic shear strain rate, $\dot{\bar{\gamma}}^p_{12}$, profiles with respect to the maximum plastic shear strain rate over the model, $\dot{\gamma}^p_{12,max}$, at the end of the analysis for different mesh discretization, for the material parameters set D1. The response converges to a mesh independent solution as the number of elements increases. The results agree well with the material length defined by \cite{DeBorst2020}. In both plots a detail near the region of interest $0.1$ m is plotted. The response was plotted without averaging between the mesh nodes (i.e. the stress computed at the Gauss points is shown).}
    \label{fig: de_borst_mesh_independent_parameters}
\end{figure}

\noindent  To investigate whether an analysis is mesh dependent or not we plot the profiles over the length of the model of the engineering total shear strain $\gamma_{12}$ and shear plastic strain rate $\dot{\gamma}^p_{12}$ for different number of elements. Results are given at the end of the analysis for time $t=0.5$ s. \\
\newline
\noindent From Figure \ref{fig: de_borst_mesh_independent_parameters},  we establish that the solution obtained with the parameters of D1 \cite{DeBorst2020} is mesh independent as the total strain $\gamma_{12}$ and the plastic shear strain rate $\dot{\gamma}^p_{12}$ profiles are spread over the model length and converge upon mesh refinement. From the analysis of \cite{DeBorst2020} we can extract the so called ``material length scale'' of the problem, which is equal to 
$l=\frac{2\eta^{vp} c}{\sqrt{\rho G}}=0.4$ m,
(based on a yield function and Perzyna material law of the form $F=\tau_{12}-c-h\gamma^p_{12},\; \dot{\gamma}^{p}_{12}=\dot{\lambda}=\frac{F}{\eta^{vp} c}$ respectively).
This is in close agreement with the total strain and plastic strain profiles we present in Figures \ref{fig: de_borst_mesh_independent_parameters}. It should be noted, however, that the above relation of \cite{DeBorst2020} does not take into account the softening slope of the material. To this end the relation available in \cite{Wang1996} can be used instead. \\

\begin{figure}[h]
    \centering
    \begin{minipage}{.37\textwidth}
    \centering
    \includegraphics[width=0.9\linewidth]{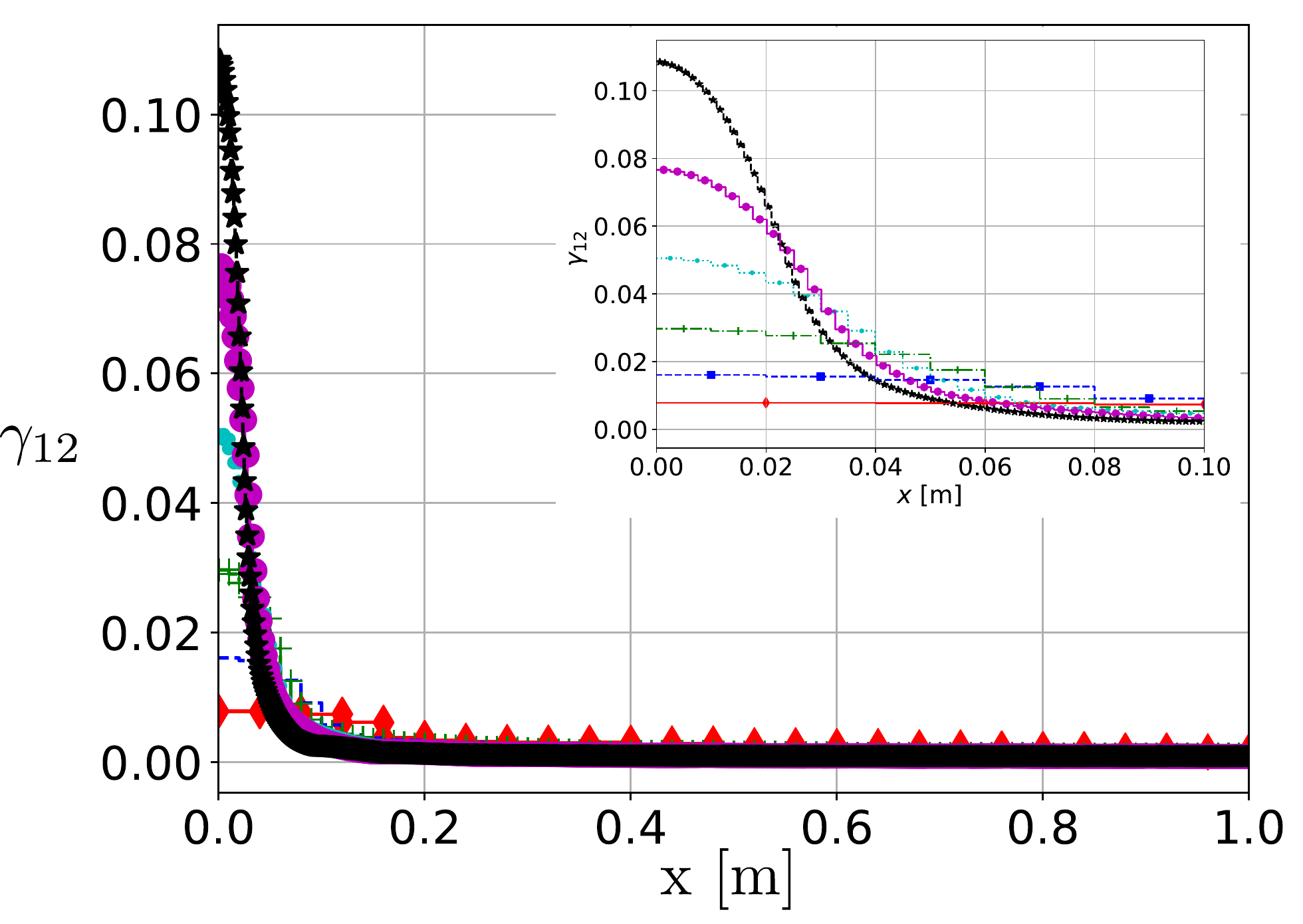}
    \end{minipage}\qquad
    \begin{minipage}{.53\textwidth}
    \centering
    \includegraphics[width=0.9\linewidth]{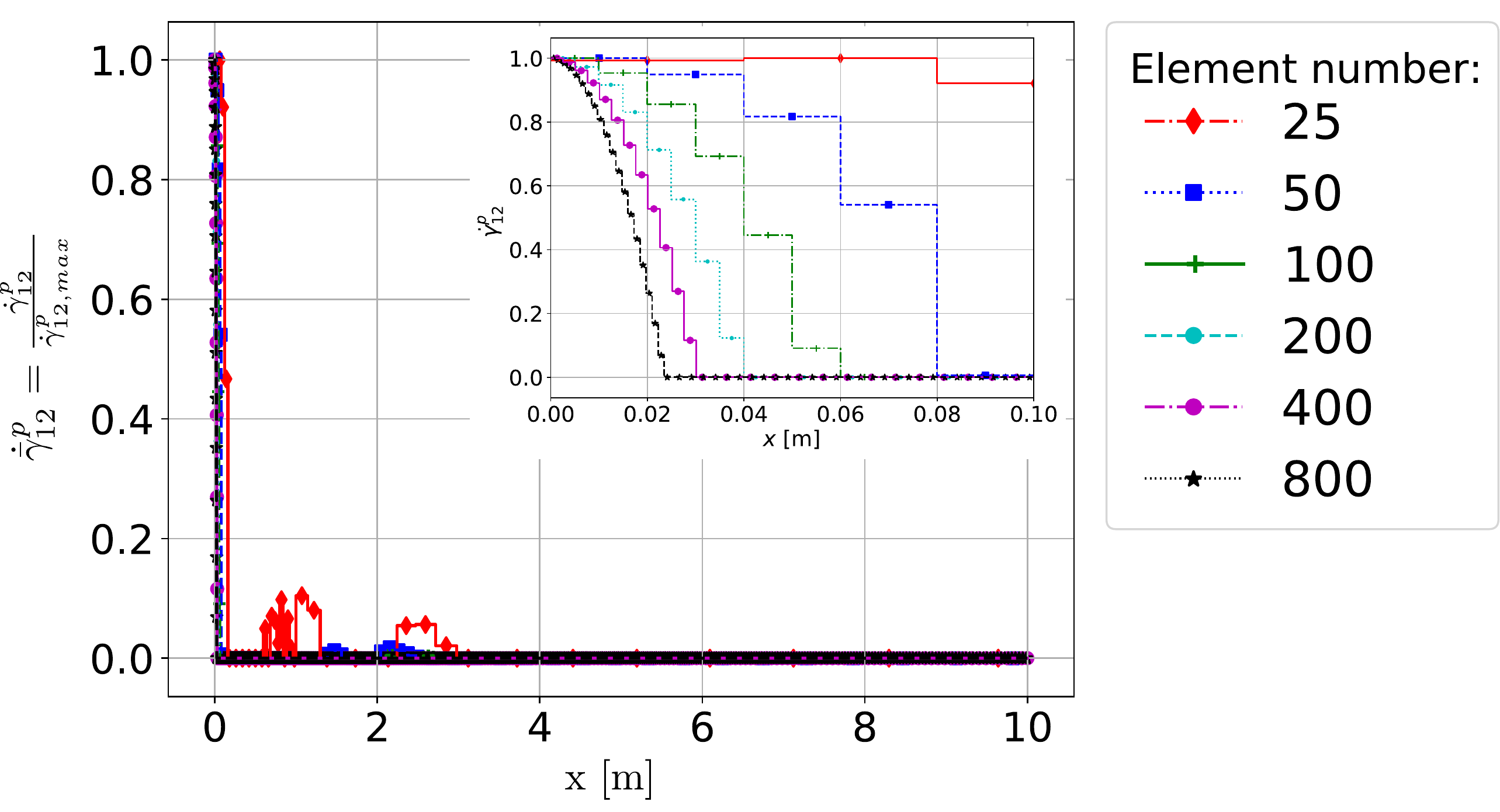}
    \end{minipage}%
    \caption{ On the left: Total shear strain $\gamma_{12}$ profiles at time  equal to 4s for different mesh discretizations of the model for the material set D1 (increased length to 10 m). Mesh dependence of the solution is observed. On the right: Normalized plastic shear strain rate, $\dot{\bar{\gamma}}^p_{12}$, profiles with respect to the maximum plastic shear strain rate over the model, $\dot{\gamma}^p_{12,max}$, at time t equal to 4 s for different mesh discretizations, for the material set D1 (increased length to 10 m). The response localizes to a mesh dependent solution as the number of elements increases. In both plots a detail near the region of interest $0.1$ m from the support is plotted. The response was plotted without averaging between the mesh nodes (i.e total strain and plastic strain rate are computed at the Gauss points).}
    \label{fig: deborst_mesh_dependent_parameters}
\end{figure}

\begin{figure}[h]
    \centering
    \begin{minipage}{.37\textwidth}
    \centering
    \includegraphics[width=0.9\linewidth]{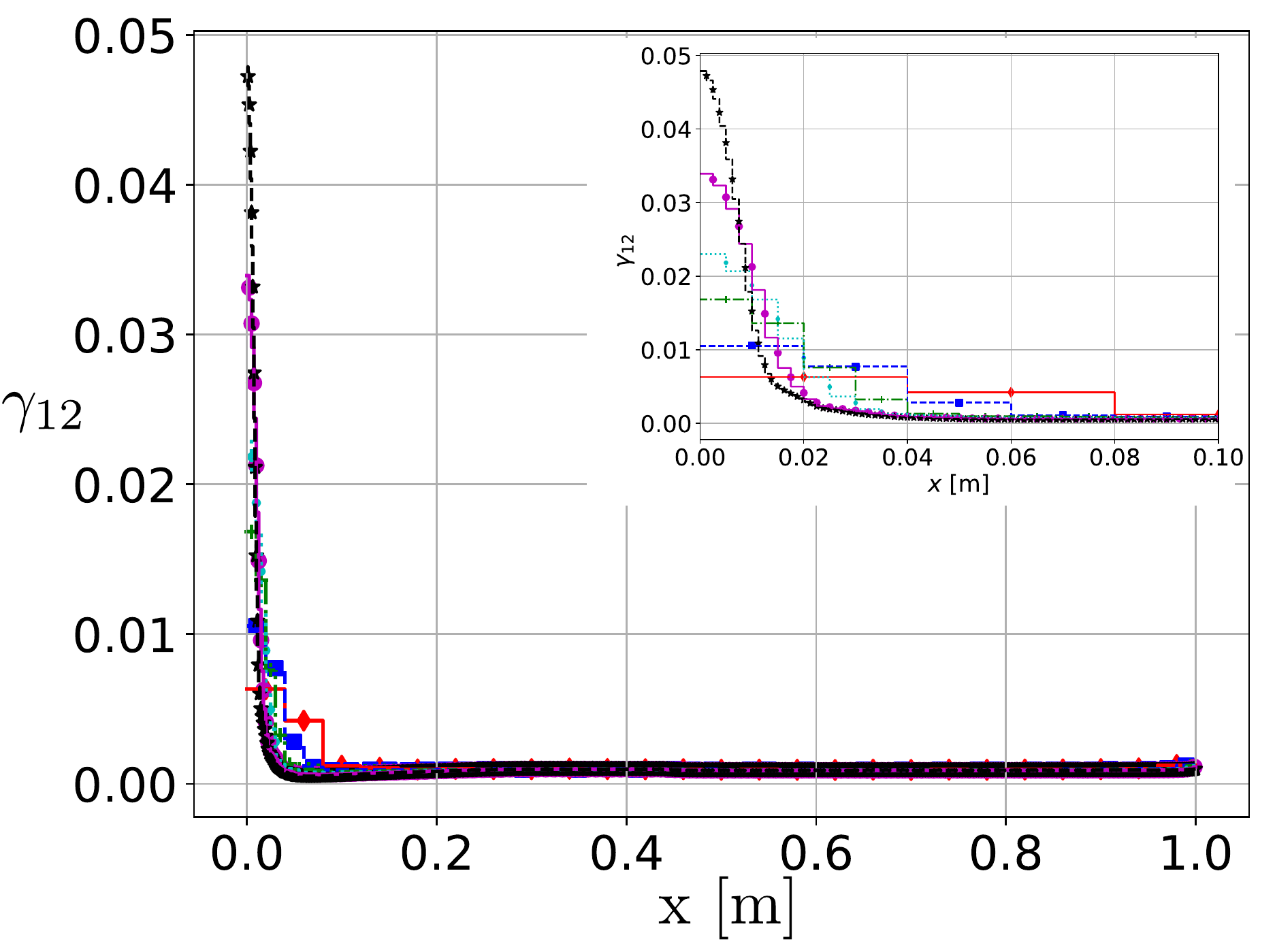}
    \end{minipage}\qquad
    \begin{minipage}{.53\textwidth}
    \centering
    \includegraphics[width=0.9\linewidth]{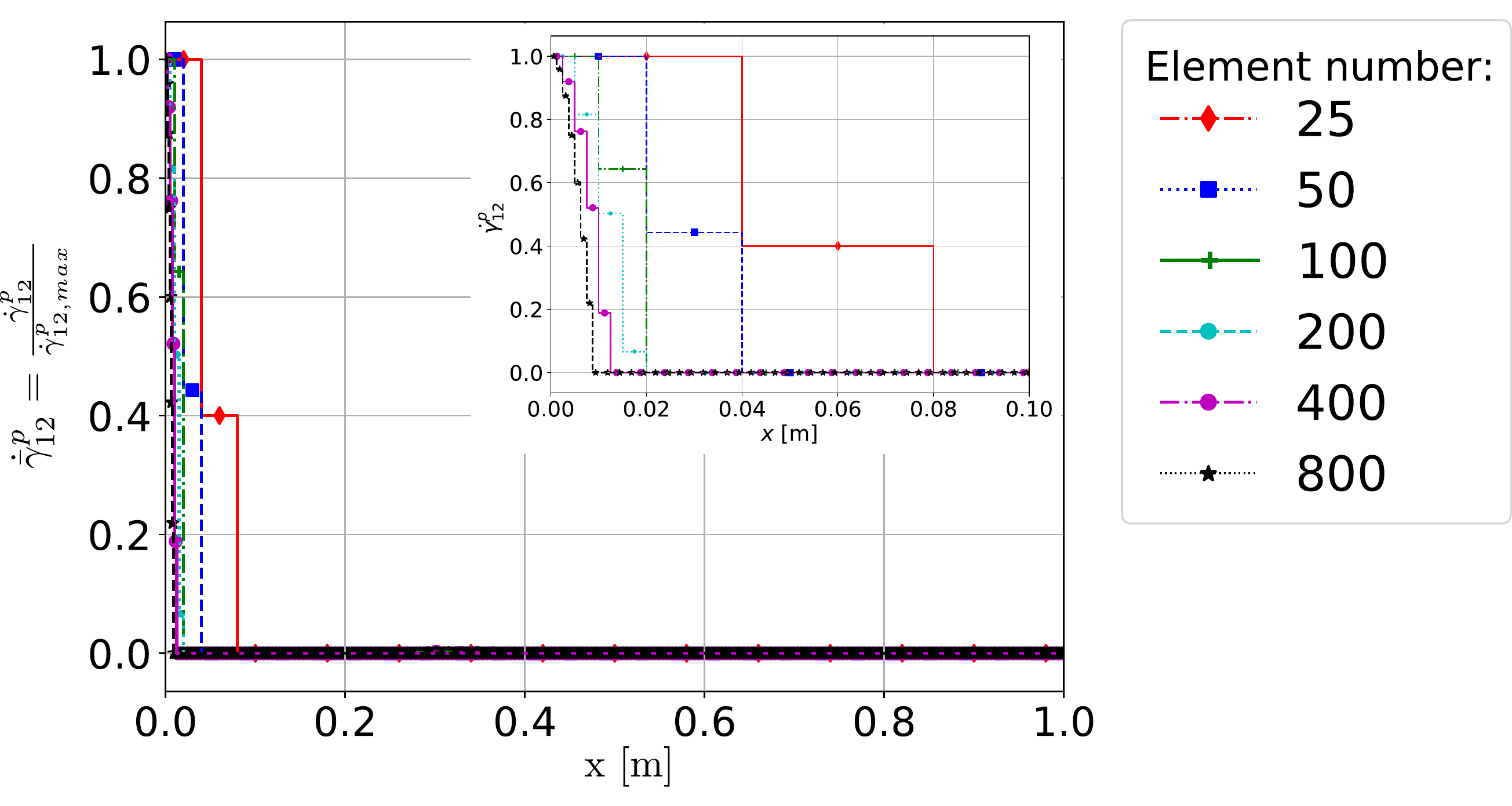}
    \end{minipage}%
    \caption{ On the left: Total shear strain $\gamma_{12}$ profiles at time equal to 0.36s for different mesh discretization of the model for the material set D2. Mesh dependence of the solution is observed. 
    On the right: Normalized plastic shear strain rate, $\dot{\bar{\gamma}}^p_{12}$, profiles with respect to the maximum plastic shear strain rate over the model, $\dot{\gamma}^p_{12,max}$, at time t equal to 0.36 s for different mesh discretizations, for the material set D2. The response localizes to a mesh dependent solution as the number of elements increases. In both plots a detail near the region of interest $0.1$ m from the edge of the support is plotted. The response was plotted without averaging between the mesh nodes (i.e total strain and plastic strain rate are computed at the Gauss points).}
    \label{fig: dalex_mesh_dependent_parameters}
\end{figure}

\noindent However, for the same material parameters of configuration D1 but for L=10 m the response of the model is completely different. We present the profiles of total shear strain and normalized plastic shear strain $\gamma_{12},\dot{\bar{\gamma}}^p_{12}$ in Figure \ref{fig: deborst_mesh_dependent_parameters}. The results are taken at time $t=4.0$ s after the stress pulse is reflected and has passed the middle of the bar for a second time. To maintain a constant mesh density near the base of the cantilever, where the highest strain gradients are observed, we apply the same uniform mesh as in the analyses of 1 m, in the 1 m region close to the cantilever support. We vary then the mesh in the rest of the cantilever by progressively increasing the element size to reduce calculation cost. Again the time increment of the analyses is kept smaller than the one specified by the CFL criterion for the smallest elements in the mesh (see Table \ref{Table: time_increment_values}). We notice that strain and plastic strain localize on one element and the solution does not show any signs of converging upon mesh refinement. The profiles show narrower localization for finer discretizations.\\
\newline
\noindent Therefore, according to the above results we can conclude that the analyses with the material set parameters D1 constitute a counterexample, about the beneficial role of viscous regularization in strain localization and mesh dependency. We emphasize here that the only thing that changed in the analysis is the length of the specimen from 1 m to 10 m.  
Finally, we present in Figure \ref{fig: dalex_mesh_dependent_parameters} another set of material parameters ( see Table \ref{Table: material_params2},  D2) that lead again to a mesh dependent behavior. The results are taken at time $t=0.36$ s after the pulse of initial stress is reflected and has passed the middle of the bar for a second time. Mesh dependence is again observed. Thus we have shown with two different counter examples (increase of the specimen's length and change of the material parameters) that viscous regularization does not lead to mesh independent results.

\subsection{Discussion}  

\noindent The difference observed in localization behavior of the analyses can be explained with the use of the theoretical findings of section \ref{sec: three}. For the material parameters of the configuration D1 the position of the pole lies in $\bar{\omega}^{P1}=-\frac{\bar{h}}{\bar{T}}=0.2t_c$, which means that the frequency of amplitude increase is equal to 0.2 $s^{-1}$ (see also equation \eqref{ubar_expanded}, Figure \ref{fig: dispersion_kr_ki} and section  \ref{sec: Linearized model: Model description }). This corresponds to a characteristic time of $T=5$ s, much larger than the time of 0.5 s which is required fro reaching steady state when $L=1$ m. If we keep the material parameters the same and increase the length of the specimen we increase the duration of the simulation without changing the dynamic character of the analysis. We provide, therefore, time to the increasing perturbation to grow. This is precisely what happens when $L=10$ m, where strain localization has time to grow enough and appear in the numerical results. The parameter set D2 corresponds to a characteristic frequency increase of the amplitude of 2 $s^{-1}$ leading to a characteristic time of 0.5 s. This is roughly equal to the total analysis time for the length of 1 m. Therefore in case D2, localization has more than enough time to develop and becomes visible before the system reaches its steady state. It must be stated here that  according to our calculations viscoplasticity can be shown to not regularize the problem irrespectively of any considerations for the magnitude of strain softening $h<0$ and strain rate hardening $g>0$ moduli. In other words based on our theoretical developments, and the counter-examples presented in this section, viscoplasticity does not regularize strain localization neither does it remedy mesh dependency.  

\section{Conclusions \label{sec: Conclusions}}

\noindent In this paper we investigated the regularization properties of elasto-viscoplasticity regarding strain localization and mesh dependency under the presence of inertia. Even though for quasi-static cases it is well-known that elasto-viscoplasticity of Perzyna or consistency type do not regularize strain localization (\cite{needleman1988material,Sluys1992}) in the dynamical case the situation was not clear.\\
\newline
\noindent Our approach is both theoretical and numerical. After deriving the equilibrium equation of the model under strain-softening $(h<0)$, strain-rate hardening $(g>0)$ elasto-viscoplasticity, we study the Lyapunov stability of states of uniform/homogeneous deformation. In order to avoid unnecessary complexity we focus on a 1D shearing example. Our mathematical analysis differs from previous ones ( see \cite{DeBorst2020,wang1997viscoplasticity,Sluys1992,needleman1988material}) by considering the frequency $\bar{\omega}$ to be a complex number in addition to the complex wavenumber $\bar{k}$. This is an important point for investigating the stability of the homogeneous, reference state as it enables the study of perturbations that can grow with time (see also  \cite{Rice1976new,lemaitre2020mecanique}).\\
\newline
\noindent Next, we proceed in finding the dispersion relationship between the complex wave number $\bar{k}$ and the complex frequency $\bar{\omega}$ for an arbitrary perturbation. The dispersion equation presents a pole, which is responsible for strain localization on a mathematical plane and thus for mesh dependency. More specifically, the wavenumber becomes infinite, $\bar{k}_r\rightarrow \infty$, and the wavelength, $\bar{\lambda}\rightarrow 0$, for strain-softening, strain-rate hardening, which means that localization on a mathematical plane is indeed possible in this system (see section \ref{sec: Characterization of the waves in the complex plane}). In the same section we have also made an extensive discussion about the possibility of traveling waves in the medium (see sections \ref{Case 2}, \ref{Case 3}) and their relation to strain localization and wave attenuation. Several qualitative observations of the poles of the dispersion equation are provided in sections \ref{Behavior_P2}, \ref{Behavior_P1}. The analysis is completed by some additional observations about the behavior of propagating sinusoidal monochromatic pulses and their relation with strain localization and phase $c_r$ and amplitude $c_i$ velocities (see Appendix \ref{sec: appendix C}).\\
\newline
\noindent We juxtapose our theoretical findings with 1D numerical analyses of an infinite layer (see section \ref{sec: Linearized model: Model description }). In particular, we investigate the effects of the perturbation mode on the localization mode. The perturbation is introduced in different shapes via various initial conditions. The theoretical relationship between the width of the perturbation and its rate of increase is confirmed. We also confirm that the smallest perturbations propagate the fastest leading to strain localization and mesh dependency. Based on these results we conclude that the elasto-viscoplastic model with strain softening and in the presence of inertia effects is unable to restrict the classical Cauchy continuum from localizing on a mathematical plane.\\
\newline
\noindent However, our analysis up to this stage, is based on a linearized version of the problem that does not take unloading into consideration. 
For this purpose, we perform fully nonlinear, dynamic numerical analyses using the ABAQUS commercial Finite Element software (\cite{0b112d0e5eba4b7f9768cfe1d818872e}) with a strain-softening, strain-rate hardening, Perzyna elasto-viscoplastic user material (UMAT). An impicit Newmark scheme was employed. Special attention was given to avoid any artificial numerical damping in order to guarantee that the right partial differential equations are solved. The results are consistent with the theoretical findings of sections \ref{sec: Bifurcation_analysis} and \ref{sec: three} and the numerical results of section \ref{sec: Linearized model: Model description } and show that mesh dependent solutions are indeed possible. It is worth noticing that for given material parameters the duration of the excitation has to be long enough in order to allow the instability to grow enough and be visible before the system reaches a steady state. This is why strain localization was not identified in previous works.\\
\newline
\noindent Our theoretical analyses show that viscoplasticity and inertia do not regularize strain localization and mesh dependency, irrespectively of the magnitude of strain softening and strain rate hardening. These results can be important for any computational method in the analysis and design of engineering products and systems in a vast variety of applications in the fields of solid mechanics, dynamics, biomechanics and geomechanics.
Our numerical analyses confirm the theoretical findings and provide counter-examples showing that viscosity and inertia do not regularize strain localization and mesh dependency.

\section*{Acknowledgments} 
\noindent The authors would like to thank Professor Nicolas Mo{\"e}s for his comments and suggestions.\\
\newline
\noindent The authors would like also to acknowledge the support of the European Research Council (ERC) under the European Union's Horizon 2020 research and innovation program (Grant agreement no. 757848 CoQuake).
\section*{Apendices}
\begin{appendices}
\numberwithin{equation}{section}
\section{Derivation of the Elasto-viscoplastic wave equation. \label{Appendix: A}}
\noindent In this Appendix the results presented in section \ref{sec: Bifurcation_analysis} are obtained in detail.

\subsection{Elasto-viscoplastic constitutive relations}
\subsubsection{Consistency model.}
\noindent In an elasto-viscoplastic formulation the following relations hold:
\begin{flalign}
&F(\sigma_{ij},\bar{\epsilon}^{vp},\dot{\bar{\epsilon}}^{vp})=0,\\
&\dot{\varepsilon}_{ij}=\dot{\varepsilon}^e_{ij}+\dot{\varepsilon}^{vp}_{ij},\\
&\dot{\sigma}_{ij}=M^e_{ijkl}\left(\dot{\varepsilon}_{kl}-\dot{\varepsilon}^{vp}_{kl}\right)\label{total_stress},\\
&\dot{\varepsilon}^{vp}_{ij}=\dot{\lambda}\frac{\partial F}{\partial \sigma_{ij}}\label{varepsilon_vp},\\
&\bar{\epsilon}^{vp} = \int^t_0\dot{\bar{\epsilon}}^{vp}dt.\\
\intertext{The viscoplastic multiplier $\dot{\lambda}$ is given by the consistency condition}
&\dot{F}=0,\;\dot{\lambda}F=0,
\end{flalign}
where $F(\sigma_{ij},\bar{\epsilon}^{vp},\dot{\bar{\epsilon}}^{vp})$ is the yield function incorporating the effects of strain and strain-rate hardening through the use of the accumulated viscoplastic strain $\bar{\epsilon}^{vp}$ and its rate $\dot{\bar{\epsilon}}^{vp}$ respectively. The time derivative of the yield condition in this case is given as:
\begin{flalign}
\dot{F}= \frac{\partial F}{\partial \sigma_{ij}}\dot{\sigma}_{ij}+\frac{\partial F}{\partial \bar{\epsilon}^{vp}}\dot{\bar{\epsilon}}^{vp}+\frac{\partial F}{\partial \dot{\bar{\epsilon}}^{vp}}\ddot{\bar{\epsilon}}^{vp}=0,
\end{flalign}
starting from the von Mises yield criterion:\\
\begin{flalign}
F(\sigma_{ij},\bar{\varepsilon}_{ij},\dot{\bar{\varepsilon}}_{ij})=\sqrt{3J_2(\sigma_{ij})}-F_0-h\bar{\epsilon}^{vp}-g\dot{\bar{\epsilon}}^{vp},
\end{flalign}
assuming dependence of yield on the deviatoric invariant of the stress tensor $J_2(s_{ij})=\sqrt{\frac{1}{2}s_{ij}s_{ij}}$, and $\dot{\bar{\epsilon}}^{vp} = \sqrt{\frac{2}{3}\dot{\varepsilon}^{vp}_{ij}\dot{\varepsilon}^{vp}_{ij}}$ where $s_{ij}=\sigma_{ij}-\frac{\sigma_{ii}}{3}$, we obtain that $\dot{\lambda}=\dot{\bar{\epsilon}}^{vp}_{ij}$ therefore the yield criterion as well as the consistency condition can be written as:
\begin{flalign}
&F(\sigma_{ij},\lambda,\dot{\lambda})=0,\\
&\dot{F}= \frac{\partial F}{\partial \sigma_{ij}}\dot{\sigma}_{ij}+\frac{\partial F}{\partial \lambda}\dot{\lambda}+\frac{\partial F}{\partial \dot{\lambda}}\ddot{\lambda}=0\label{consistency_condition},
\end{flalign} 
multiplying \eqref{total_stress} by $\frac{\partial F}{\partial \sigma_{ij}}$. replacing $\dot{\varepsilon}^{vp}_{ij}$ with help from \eqref{varepsilon_vp} and replacing the term in the lefthandside with equation \eqref{consistency_condition} we finally get:
\begin{flalign}
&-\frac{\partial F}{\partial \lambda}\dot{\lambda}-\frac{\partial F}{\partial \dot{\lambda}}\ddot{\lambda}=\frac{\partial F}{\partial \sigma_{ij}}M^e_{ijkl}\left(\dot{\varepsilon}_{kl}-\dot{\lambda}\frac{\partial F}{\partial \sigma_{kl}}\right).
\intertext{Grouping together the terms of $\dot{\lambda}$ and solving for $\dot{\lambda}$ we get:}
&\dot{\lambda} = \frac{\frac{\partial F}{\partial \sigma_{ij}}M^e_{ijkl}}{-\frac{\partial F}{\partial \lambda}+\frac{\partial F}{\partial \sigma_{ij}}M^e_{ijkl}\frac{\partial F}{\partial \sigma_{kl}}}\dot{\varepsilon}_{kl}+\frac{\frac{\partial F}{\partial \dot{\lambda}}}{-\frac{\partial F}{\partial \lambda}+\frac{\partial F}{\partial \sigma_{ij}}M^e_{ijkl}\frac{\partial F}{\partial \sigma_{kl}}}\ddot{\lambda}\label{ldot},
\intertext{Inserting \eqref{ldot} into \eqref{total_stress} we obtain:}
&\dot{\sigma}_{ij}=M^e_{ijkl}\left(\dot{\varepsilon}_{kl}-\frac{\frac{\partial F}{\partial \sigma_{ij}}M^e_{ijkl}\frac{\partial F}{\partial \sigma_{kl}}}{-\frac{\partial F}{\partial \lambda}+\frac{\partial F}{\partial \sigma_{ij}}M^e_{ijkl}\frac{\partial F}{\partial \sigma_{kl}}}\dot{\varepsilon}_{kl}+\frac{\frac{\partial F}{\partial \dot{\lambda}}\frac{\partial F}{\partial \sigma_{kl}}}{-\frac{\partial F}{\partial \lambda}+\frac{\partial F}{\partial \sigma_{ij}}M^e_{ijkl}\frac{\partial F}{\partial \sigma_{kl}}}\ddot{\lambda}\right),
\intertext{Replacing the time derivative with a variation taking advantage of the definition of variation we arrive at the constitutive equation describing the perturbed field of stress $\tilde{\sigma}_{ij}$.}
&\tilde{\sigma}_{ij}=M^e_{ijkl}\left(\tilde{\varepsilon}_{kl}-\frac{\frac{\partial F}{\partial \sigma_{ij}}M^e_{ijkl}\frac{\partial F}{\partial \sigma_{kl}}}{-\frac{\partial F}{\partial \lambda}+\frac{\partial F}{\partial \sigma_{ij}}M^e_{ijkl}\frac{\partial F}{\partial \sigma_{kl}}}\tilde{\varepsilon}_{kl}+\frac{\frac{\partial F}{\partial \dot{\lambda}}\frac{\partial F}{\partial \sigma_{kl}}}{-\frac{\partial F}{\partial \lambda}+\frac{\partial F}{\partial \sigma_{ij}}M^e_{ijkl}\frac{\partial F}{\partial \sigma_{kl}}}\dot{\tilde{\lambda}}\right)\label{constitutive_law_perturbed}.
\end{flalign}
\subsection{Derivation of the perturbed equation\label{sec:perturbed_eq}}
\noindent Inserting \eqref{constitutive_law_perturbed} into \eqref{perturbed_equation} we arrive at:
\begin{flalign}
&M^e_{ijkl}\left(\tilde{\varepsilon}_{kl,j}-\frac{\frac{\partial F}{\partial \sigma_{ij}}M^e_{ijkl}\frac{\partial F}{\partial \sigma_{kl}}}{-\frac{\partial F}{\partial \lambda}+\frac{\partial F}{\partial \sigma_{ij}}M^e_{ijkl}\frac{\partial F}{\partial \sigma_{kl}}}\tilde{\varepsilon}_{kl,j}+\frac{\frac{\partial F}{\partial \dot{\lambda}}\frac{\partial F}{\partial \sigma_{kl}}}{-\frac{\partial F}{\partial \lambda}+\frac{\partial F}{\partial \sigma_{ij}}M^e_{ijkl}\frac{\partial F}{\partial \sigma_{kl}}}\dot{\tilde{\lambda}}_{,j}\right)=\rho \ddot{\tilde{u}}_{i}\label{eq_perturbed_1}.\\
\intertext{Substituting \eqref{varepsilon_vp} into the last term of the right hand side of eq. \eqref{eq_perturbed_1}  such that $\dot{\varepsilon}^{vp}_{kl}=\frac{\partial F}{\partial \sigma_{kl}}\dot{\lambda}$ we arrive at:} 
&M^e_{ijkl}\left(\tilde{\varepsilon}_{kl,j}-\frac{\frac{\partial F}{\partial \sigma_{ij}}M^e_{ijkl}\frac{\partial F}{\partial \sigma_{kl}}}{-\frac{\partial F}{\partial \lambda}+\frac{\partial F}{\partial \sigma_{ij}}M^e_{ijkl}\frac{\partial F}{\partial \sigma_{kl}}}\tilde{\varepsilon}_{kl,j}+\frac{\frac{\partial F}{\partial \dot{\lambda}}}{-\frac{\partial F}{\partial \lambda}+\frac{\partial F}{\partial \sigma_{ij}}M^e_{ijkl}\frac{\partial F}{\partial \sigma_{kl}}}\dot{\tilde{\varepsilon}}^{vp}_{kl,j}\right)=\rho \ddot{\tilde{u}}_{i},
\intertext{rewritting $\dot{\tilde{\varepsilon}}^{vp}_{kl}=\dot{\tilde{\varepsilon}}_{kl}-{M^{e^{{-1}}}_{ijkl}}\dot{\tilde{\sigma}}_{ij}$ :}
&M^e_{ijkl}\left(\tilde{\varepsilon}_{kl,j}-\frac{\frac{\partial F}{\partial \sigma_{ij}}M^e_{ijkl}\frac{\partial F}{\partial \sigma_{kl}}}{-\frac{\partial F}{\partial \lambda}+\frac{\partial F}{\partial \sigma_{ij}}M^e_{ijkl}\frac{\partial F}{\partial \sigma_{kl}}}\tilde{\varepsilon}_{kl,j}+\frac{\frac{\partial F}{\partial \dot{\lambda}}}{-\frac{\partial F}{\partial \lambda}+\frac{\partial F}{\partial \sigma_{ij}}M^e_{ijkl}\frac{\partial F}{\partial \sigma_{kl}}}\dot{\tilde{\varepsilon}}_{kl,j}-{M^{e^{{-1}}}_{ijkl}}\dot{\tilde{\sigma}}_{ij,j}\right)=\rho \ddot{\tilde{u}}_{i},\\
\intertext{inserting finally \eqref{perturbed_equation} we obtain:}
&M^e_{ijkl}\left(\tilde{\varepsilon}_{kl,j}-\frac{\frac{\partial F}{\partial \sigma_{ij}}M^e_{ijkl}\frac{\partial F}{\partial \sigma_{kl}}}{-\frac{\partial F}{\partial \lambda}+\frac{\partial F}{\partial \sigma_{ij}}M^e_{ijkl}\frac{\partial F}{\partial \sigma_{kl}}}\tilde{\varepsilon}_{kl,j}+\frac{\frac{\partial F}{\partial \dot{\lambda}}}{-\frac{\partial F}{\partial \lambda}+\frac{\partial F}{\partial \sigma_{ij}}M^e_{ijkl}\frac{\partial F}{\partial \sigma_{kl}}}\dot{\tilde{\varepsilon}}_{kl,j}-{M^{e^{{-1}}}_{ijkl}}\rho\dddot{\tilde{u}}_{i}\right)=\rho \ddot{\tilde{u}}_{i}.
\end{flalign}
\subsubsection{1D Example: shearing of a viscous Cauchy layer}
\noindent We proceed in deriving the perturbed linear momentum equation for the shearing of a 1D elasto-visoplastic Cauchy layer. The basic kinematic equations and the constitutive relations with the von Mises yield criterion with linear strain and strain-rate hardening/softening, are given as follows:
\begin{flalign}
&F(\tau,\lambda) = \sqrt{3}\tau-\bar{\tau}({\lambda})-\eta F_0\dot{\lambda},\\
&\dot{\gamma} = \dot{\gamma}^{el}+\dot{\gamma}^{vp},\\
&\dot{\gamma}^{vp} = \dot{\lambda}\frac{\partial F}{\partial \tau}=\frac{F}{\eta F_{0}}\frac{\partial F}{\partial \tau},\\
&\bar{\tau}({\lambda}) =F_0 + H{\lambda} = F_0 + \frac{H}{\sqrt{3}}\gamma^{vp},
\label{Perzyna_model}
\end{flalign}
where $\tau$ is the shear stress, $\gamma=2\varepsilon_{12}=2\varepsilon_{21}=u_{2,1}$ is the engineering shear strain, $F_0$ is the initial yield strength. $h$ is a hardening/softening material parameters with units of pressure (MPa), while $\eta$ is the viscosity parameter with units of time $s$.\\
Applying the procedure described above and taking advantage of the viscosity potential $F(\tau,\lambda,\dot{\lambda})$ assuming the consistency condition $\dot{F} = 0$, we obtain the following:
\begin{flalign}
&\dot{F}=\frac{\partial F}{\partial \tau}\dot{\tau} + \frac{\partial F}{\partial \lambda}\dot{\lambda}+ \frac{\partial F}{\partial \dot{\lambda}}\ddot{\lambda}=\frac{\partial F}{\partial \tau}\dot{\tau}-\frac{\partial \bar{\tau}}{\partial \lambda}\dot{\lambda}-\eta F_{0}\ddot{\lambda},\\
&\dot{\tau} = G\left(\dot{\gamma}-\dot{\gamma}^{vp}\right)\label{elastic_stress_rate}.\\
\intertext{Multiplying the above equation \eqref{elastic_stress_rate} with $\frac{\partial F}{\partial \tau} $ and replacing $\frac{\partial F}{\partial \tau}\dot{\tau}$ we get:}
&\dot{\tau}\sqrt{3} = G\left(\dot{\gamma}-\dot{\gamma}^{vp}\right)\sqrt{3},\\
&\eta F_{0}\ddot{\lambda}+\frac{\partial \bar{\tau}}{\partial \lambda}\dot{\lambda} = G\left(\dot{\gamma}-\dot{\gamma}^{vp}\right)\sqrt{3},\\
\intertext{separating $\dot{\lambda}$ we obtain:}
&\dot{\lambda} = \frac{G \dot{\gamma} \sqrt{3}}{\frac{\partial \bar{\tau}}{\partial \lambda}+3G}+\frac{\eta F_{0} \ddot{\lambda}}{\frac{\partial \bar{\tau}}{\partial \lambda}+3G},\\
\intertext{substituting $\dot{\lambda}$ to the original equation for the calculation of $\dot{\tau}$:}
&\dot{\tau} = G\left(\dot{\gamma}-\frac{3G}{\frac{\partial \bar{\tau}}{\partial \lambda}+3G}\right) +\frac{G\eta F_{0}\sqrt{3}\ddot{\lambda}}{\frac{\partial \bar{\tau}}{\partial \lambda}+3},\\
\intertext{substituting $\frac{1}{G}\frac{\partial \bar{\tau}}{\partial \lambda}=\frac{H}{G}=\bar{h}$ (linear mechanical softening) we arrive finally at:}
\dot{\tau} &= G\left(\frac{\bar{h}}{\bar{h}+3}\right)\dot{\gamma} +\frac{\eta F_{0}}{\bar{h}+3}\ddot{\gamma}^{vp}\label{constitutive_relation}.\\
\intertext{It should be noted that the results derived until now can be also obtained using the Perzyna model instead of the consistency approach in the case of monotonic loading (no stress reversal, so that the rate dependent unloading overstress of the Perzyna model is not taken into account). Using the Perzyna material we assume the existence of the viscoplastic potential $\Omega(\tau,\lambda,\dot{\lambda})$ which constitutes a region outside the yield function that the stress vector $\tau$ is indeed applicable. The time derivative of the yield function $\dot{F}$ is the given as:} 
&\dot{F}=\eta F_0 \ddot{\lambda} = \frac{\partial F}{\partial \tau}\dot{\tau}+\frac{\partial F}{\partial \lambda}\dot{\lambda}=\frac{\partial F}{\partial \tau}\dot{\tau}-\frac{\partial \bar{\tau}}{\partial \lambda}\dot{\lambda}\label{consistency_cond_2}.\\
\intertext{Using the same arguments as before, (multiplying \eqref{elastic_stress_rate} by $\frac{\partial F}{\partial \tau}$ and substituting \eqref{consistency_cond_2}) we arrive at the same expression for $\dot{\lambda}$ and $\dot{\tau}$.}
\intertext{Perturbing and replacing \eqref{constitutive_relation} in equation \eqref{perturbed_equation} we obtain:}
&G\frac{\bar{h}}{3+\bar{h}}\frac{\partial \tilde{\gamma}}{\partial x}+\frac{\eta F_0}{3+\bar{h}}\frac{\partial \dot{\tilde{\gamma}}^{vp}}{\partial x}=\rho\ddot{\tilde{u}},\\
&G\frac{\bar{h}}{3+\bar{h}}\frac{\partial \tilde{\gamma}}{\partial x}+\frac{\eta F_0}{3+\bar{h}}\left(\frac{\partial \dot{\tilde{\gamma}}}{\partial x}-\frac{1}{G}\frac{\partial \dot{\tilde{\tau}}}{\partial x}\right)=\rho\ddot{\tilde{u}},\\
&G\frac{\bar{h}}{3+\bar{h}}\frac{\partial^2 \tilde{u}}{\partial x^2}-\rho\frac{\partial^2 \tilde{u}}{\partial t^2}+\frac{\eta F_0}{(3+\bar{h})}\left[\frac{\partial^3 \tilde{u}}{\partial t \partial x^2}-\frac{1}{v^2_s}\frac{\partial^3 \tilde{u}}{\partial t^3}\right]=0,\\
&G\bar{h}\frac{\partial^2 \tilde{u}}{\partial x^2}-\frac{\partial^2 \tilde{u}}{\partial t^2}\frac{(3+\bar{h})G}{v^2_s}+\bar{\eta}^{vp}G\left[\frac{\partial^3 \tilde{u}}{\partial t \partial x^2}-\frac{1}{v^2_s}\frac{\partial^3 \tilde{u}}{\partial t^3}\right]=0\label{viscoelastic_wave},
\end{flalign}
where $v_s=\sqrt{\frac{G}{\rho}},\;\bar{\eta}^{vp}G=\eta F_0$.
\subsubsection{Normalizing the 1D elasto-viscoplastic wave equation.}
\noindent We consider $\bar{u}=\frac{u}{u_c},\bar{t}=\frac{t}{t_c},\bar{x}=\frac{x}{x_c}$. Applying and differentiating into \eqref{viscoelastic_wave} we arrive at:
\begin{flalign}
&\frac{x^2_c}{v^2_s t^2_c}\frac{\partial^3\bar{u}}{\partial \bar{t}^3}-\frac{\partial^3\bar{u}}{\partial \bar{x}^2\partial \bar{t}}\frac{\bar{\eta}^{vp}}{t_c \bar{h}}+\frac{x^2_c}{v^2_s t^2_c}\frac{3+\bar{h}}{\bar{h}}\frac{\partial^2 \bar{u}}{\partial \bar{t}^2}-\frac{\partial^2 \bar{u}}{\partial \bar{x}^2}=0.
\intertext{Introducing the characteristic velocity $v_c=\frac{x_c}{t_c}$, the result is written as:}
&\frac{v^2_c}{v^2_s }\frac{\partial^3\bar{u}}{\partial \bar{t}^3}-\frac{\partial^3\bar{u}}{\partial \bar{x}^2\partial \bar{t}}\frac{\bar{\eta}^{vp}}{t_c \bar{h}}+\frac{v^2_c}{v^2_s}\frac{3+\bar{h}}{\bar{h}}\frac{\partial^2 \bar{u}}{\partial \bar{t}^2}-\frac{\partial^2 \bar{u}}{\partial \bar{x}^2}=0\label{normalized equation}.
\end{flalign}
\subsubsection{Dispersion relationship}
\noindent Inserting into the normalized equation \eqref{normalized equation} the nondimensional solution assuming both $\bar{k},\bar{\omega} \;\in \; \mathds{C}$ :
\begin{flalign}
\bar{u}(\bar{x},\bar{t}) = \exp{i( \bar{k} \bar{x}-\bar{\omega} \bar{t})},
\end{flalign}
we arrive at:
\begin{flalign}
&\bar{h} \bar{k}^2 t_c v^2_s - i \bar{k}^2 v^2_s \bar{\eta}^{vp} \bar{\omega} - (3 + \bar{h}) \bar{t_c} v^2_c \bar{\omega}^2 + i v^2_c \bar{\eta}^{vp}\bar{\omega}^3 = 0\label{dispersion_relation}.
\end{flalign}


\section{Behavior of the dispersion relation near infinity.}
\subsection{Properties of the inverse transform $\bar{\omega}=\frac{1}{z}$ \label{Appendix_B}}

\noindent We present here in more detail the inverse transform $\bar{\omega}=\frac{1}{z}$ based on \cite{brown2009complex}. The inverse transform allows the analysis of the behavior of a function close to $\infty$ by inverting the independent variable. Therefore, as the independent variable $\bar{\omega}$ tends to $\infty$, its inverse transform $z$ tends to 0. More specifically the points lying outside the unit circle centered at the origin of the $\bar{\omega}$ complex plane are mapped inside the circle, while the opposite is true for the points initially inside the unit circle. The points lying on the unit circle remain on the circle.\\
\noindent When a point $\bar{\omega}=\bar{\omega}_r+\bar{\omega}_i i$ is the image of a nonzero point $z=z_r+z_i i$ under the transformation $\bar{\omega}=1/z$, the relationship between the real and imaginary parts in original $\bar{\omega}$ and transformed $z$ complex planes respectively are given as:
\begin{align}
z_r=\frac{\bar{\omega}_r}{\bar{\omega}^2_r+\bar{\omega}^2_i}\;\text{and}\;z_i=-\frac{\bar{\omega}_i}{\bar{\omega}^2_r+\bar{\omega}^2_i},\\
\bar{\omega}_r=\frac{z_r}{z^2_r+z^2_i}\;\text{and}\;\bar{\omega}_i=-\frac{z_i}{z^2_r+z_i}.
\end{align}

\noindent When A,B,C,D $\in \mathbb{R}$ are real numbers satisfying the condition $\text{B}^2+\text{C}^2>4\text{A}\text{D}$, the equation
\begin{align}
\text{A}(z^2_r+z^2_i)+\text{B}z_r+\text{C}z_i+\text{D}=0 \label{circle-line_in_z}
\end{align} 

\noindent represents a circle or a line on the complex plane z. In particular, when $\text{A}= 0$ the equation of a line is returned while when $\text{A}\neq 0$ by completing the squares we get:
\begin{align}
\left(z_r+\frac{\text{B}}{\text{2A}}\right)^2+\left(z_i+\frac{\text{C}}{\text{2A}}\right)^2=\left(\frac{\sqrt{\text{B}^2+\text{C}^2-4\text{A}\text{D}}}{2\text{A}}\right).
\end{align}
\noindent The above equation represents a circle under the condition mentioned previously. Similarly by substitution of $z_r,z_i$ we find: 
\begin{align}
\text{D}(\bar{\omega}^2_r+\bar{\omega}^2_i)+\text{B}\bar{\omega}_r-\text{C}\bar{\omega}_i+\text{A}=0\label{circle_line_in_omega}.
\end{align} 
\noindent From the equations \eqref{circle-line_in_z},\eqref{circle_line_in_omega} above it is clear that:
\begin{itemize}
\item A circle in the $z$-plane ($\text{A}\neq 0$) not passing through the origin ($\text{D}\neq 0$) is transformed into a circle not passing through the origin in the $\bar{\omega}$-plane.
\item A circle in the $z$-plane ($\text{A}\neq 0$) passing through the origin ($\text{D}=0$) is transformed into a line not passing through the origin in the $\bar{\omega}$-plane.
\item A line in the $z$-plane ($\text{A}= 0$) not passing through the origin ($\text{D}\neq 0$) is transformed into a circle passing through the origin in the $\bar{\omega}$-plane.
\item A line in the $z$-plane ($\text{A}\neq 0$) passing through the origin ($\text{D}=0$) is transformed into a line passing through the origin in the $\bar{\omega}$-plane. 
\end{itemize}

\noindent From the above remarks we conclude that in the two complex planes the directions of the real and imaginary axes coincide. Furthermore every line passing from the origin retains its direction.\\
\newline
\noindent For our analyses we need to examine the behavior of $\bar{k}_{1,2}(\bar{\omega})$ along lines of constant $\bar{\omega}_r=c_1,\bar{\omega}_i=c_2$. The geometrical loci on the complex $z$-plane are given by equations:

\begin{align}
\left(z_r-\frac{1}{2c_1}\right)^2+z^2_i=\left(\frac{1}{c_1}\right)^2,\\
z^2_r+\left(z_i+\frac{1}{2c_2}\right)^2=\left(\frac{1}{c_2}\right)^2.
\end{align}  

\noindent We notice that due to the inverse transform properties, lines parallel to the real axis $\text{Re}$, lying in one half of the complex $\bar{\omega}$-plane are transformed into circles passing through the origin, whose center lies on the opposite imaginary half of the $z$-plane.\\
\subsection{Application of the inverse mapping describing the point at complex infinity.}
\noindent Applying this mapping in equation \eqref{final_form_k1} yields:
\begin{flalign}
\bar{k}_1(z)=V\frac{1}{z}\left(\frac{3+\bar{h}}{iT}-\frac{1}{z}\right)^{\frac{1}{2}}\left(\frac{\bar{h}}{iT}-\frac{1}{z}\right)^{-\frac{1}{2}} \label{solution inverse mapping},
\end{flalign}
where the following relations hold between the components of $\bar{\omega},z$ in their respective complex plane:
\begin{align}
z_r=\frac{\bar{\omega}_r}{\bar{\omega}^2_r+\bar{\omega}^2_i}\;,\;
z_i=\frac{-\bar{\omega}_i}{\bar{\omega}^2_r+\bar{\omega}^2_i}.
\end{align}
\noindent Taking the limit as $z\rightarrow 0$ and expanding the root terms as Taylor series around $z\rightarrow 0$ we obtain:
\begin{flalign}
\lim_{z\rightarrow 0}\bar{k}_1\left(\frac{1}{z}\right)=V\frac{1}{z}\left(1-\frac{1}{2}\frac{3+\bar{h}}{iT}z+\frac{1}{8}\left(\frac{3+\bar{h}}{iT}\right)^{2}z^2+...\right)\frac{1}{1-\left(\frac{1}{2}\frac{3+\bar{h}}{iT}z-\frac{1}{8}\left(\frac{3+\bar{h}}{iT}\right)^{2}z^2+...\right)}.
\end{flalign}
\noindent We notice the pattern in the denominator of the last term that we can replace with the Taylor series of $\frac{1}{1-x}$ around $x \rightarrow 0$ leading to:
\begin{flalign}
\begin{aligned}
\lim_{z\rightarrow 0}\bar{k}_1\left(z\right)=&V\frac{1}{z}\left(1-\frac{1}{2}\frac{3+\bar{h}}{iT}z+\frac{1}{8}\left(\frac{3+\bar{h}}{iT}\right)^{2}z^2+...\right)\left(1+\left(\frac{1}{2}\frac{3+\bar{h}}{iT}z+\frac{1}{8}\left(\frac{3+\bar{h}}{iT}\right)^{2}z^2+...\right)\right. +\\
&\left. +\left(\frac{1}{2}\frac{3+\bar{h}}{iT}z+\frac{1}{8}\left(\frac{3+\bar{h}}{iT}\right)^{2}z^2+...\right)^2+...\right)\label{localization_proof}.
\end{aligned}
\end{flalign}
From the above polynomial only the factor $\frac{1}{z}$ tends to $\infty$, therefore $z=0 \Leftrightarrow \bar{\omega}^{P2}\rightarrow \infty$ is a pole of first order \cite{brown2009complex}.\\
\newline
\noindent For the real part of $\bar{k}$ at the pole $\bar{\omega}^{P2}$ as $z_r$ tends to $\infty$ applying de l' Hopital rule we can prove:

\begin{align} 
\lim_{\bar{z}_r\rightarrow 0^+}\text{Re}\left[\bar{k}_{1,2}(\bar{z}_r+\beta i)\right]=\lim_{\bar{z}_r\rightarrow 0^+}\left[\pm V1\sqrt[4]{\frac{(3+h)^2+\frac{T1^2}{z^2_r}}{{h^2+\frac{T1^2}{\bar{z}^2_r}}}}\frac{\cos{\left(\frac{1}{2}\arg{\left(\frac{3+h-\frac{iT1}{\bar{z}_r}}{h-\frac{iT1}{\bar{z}_r}}\right)}\right)}}{\bar{z}_r}\right]=\pm \infty, \label{kr_infty}
\end{align}
where $\beta=0$ in the case of traveling waves.

\noindent Furthermore, for the imaginary part  of $\bar{k}(z)$ at the pole $\bar{\omega}^{P2}$ as $z_r$ tends to $\infty$ we can prove that $\bar{k}_i \rightarrow \pm c\;\in \mathbb{R}$ using the de l' H\^{o}pital rule: 

\begin{align} 
\lim_{\bar{z}_r\rightarrow 0}\text{Im}\left[\bar{k}_{1,2}(\bar{z}_r+0i)\right]=\lim_{\bar{z}_r\rightarrow 0}\left[\pm V1\sqrt[4]{\frac{(3+h)^2+\frac{T1^2}{z^2_r}}{{h^2+\frac{T1^2}{\bar{z}^2_r}}}}\frac{\sin{\left(\frac{1}{2}\arg{\left(\frac{3+h-\frac{iT1}{\bar{z}_r}}{h-\frac{iT1}{\bar{z}_r}}\right)}\right)}}{\bar{z}_r}\right]=\pm c \label{constant k_i}
\end{align}

\noindent The value of $\bar{k}_i(\bar{\omega})$ is independent of whether we move towards the left or the right of the real axis $\bar{\omega}_r$. It only depends on the solution branch $\bar{k}_{1}(\bar{\omega}),$ or $\bar{k}_{2}(\bar{\omega})$ we follow. (see Figure \ref{fig: dispersion_ki_omegar}).

\section{Stability and localization of a monochromatic sinusoidal propagating pulse \label{sec: appendix C}}
\begin{figure}[H]
    \centering
    \begin{minipage}{.45\textwidth}
    \centering
    \includegraphics[width=0.9\textwidth]{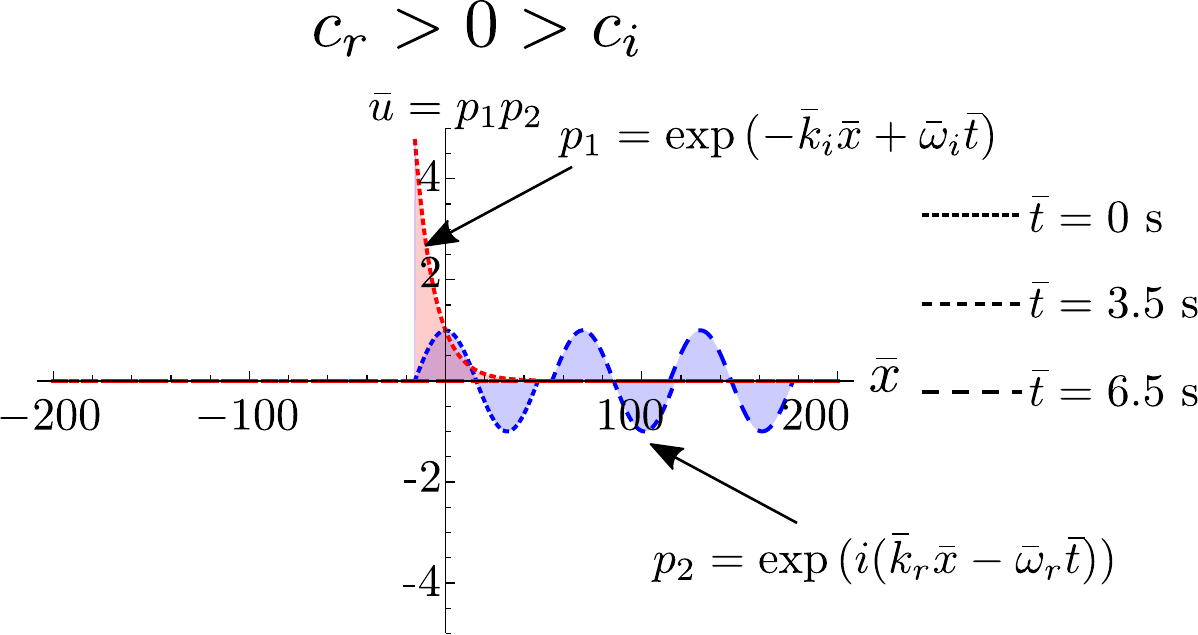}
	\end{minipage}\qquad
	\begin{minipage}{.45\textwidth}
    \centering
    \includegraphics[width=0.9\textwidth]{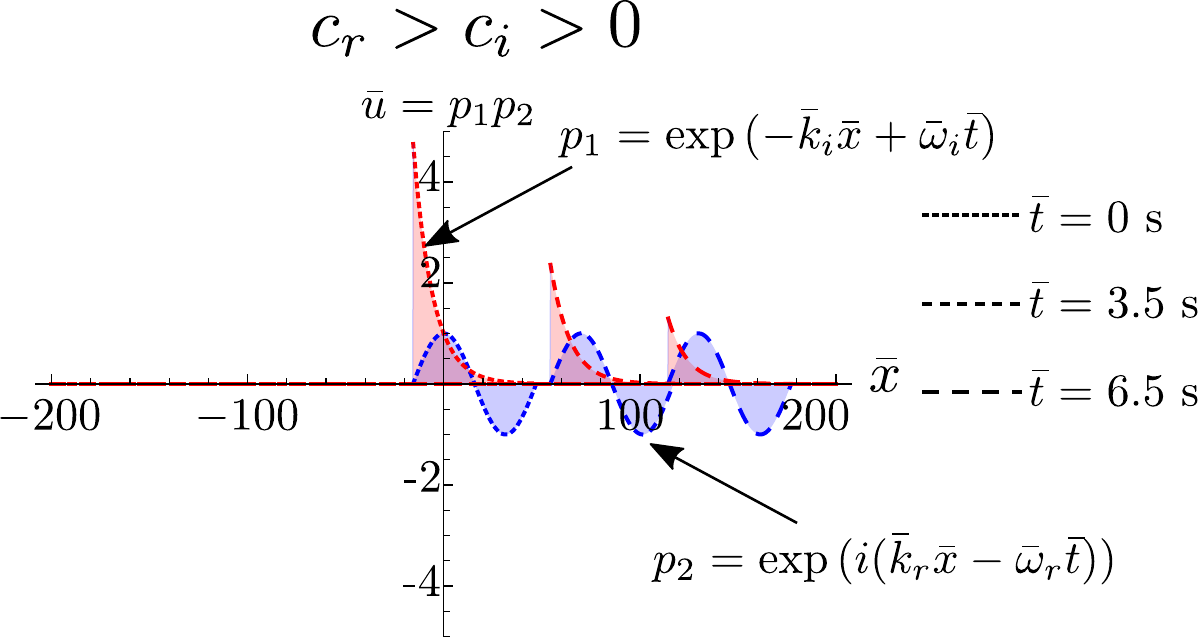}
    \end{minipage}
    \caption{Conditions for decaying perturbation. Left: Evolution of the perturbation (blue curve, $\exp{(i(\bar(x)-c_r \bar{t})}$) and its amplitude (red curve, $\exp{(-(\bar{x}-c_i\bar{t}))}$) at different times for $c_i<0<c_r$. Right: Evolution of the perturbation (blue curve) and its amplitude (red curve) at different times for $c_r>c_i>0$. The propagating pulse is the multiplication of the red and blue curves.}
    \label{fig:cr_ci_1}
\end{figure}

\begin{figure}[]
    \centering
    \begin{minipage}{.45\textwidth}
    \centering
    \includegraphics[width=0.9\textwidth]{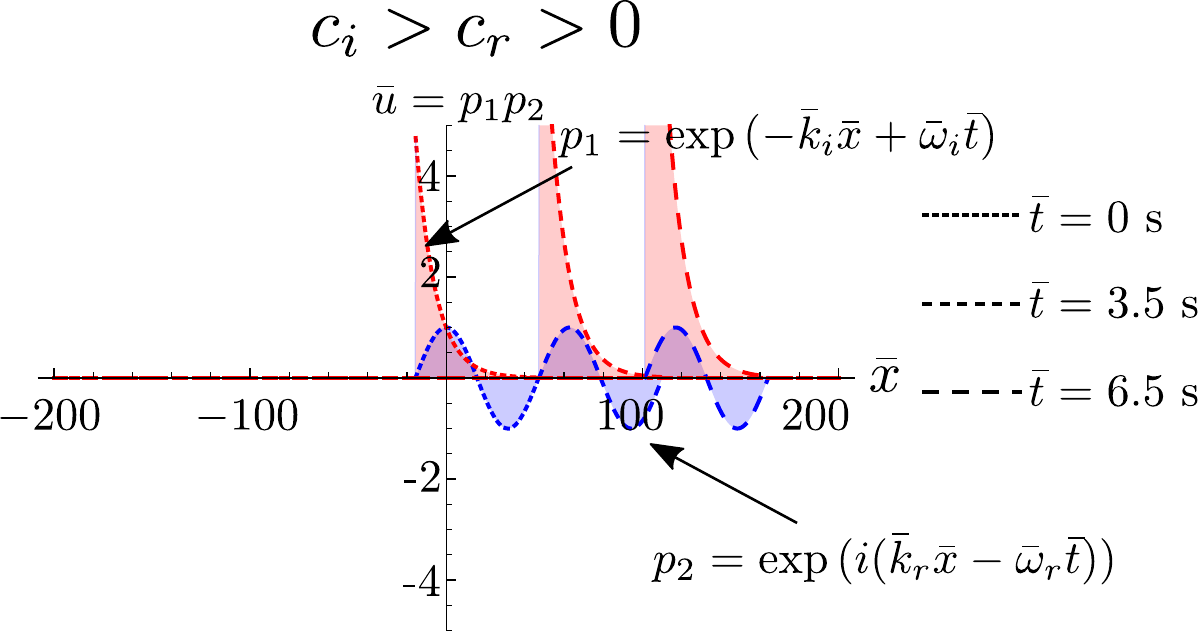}
	\end{minipage}\qquad
	\begin{minipage}{.45\textwidth}
    \centering
    \includegraphics[width=0.9\textwidth]{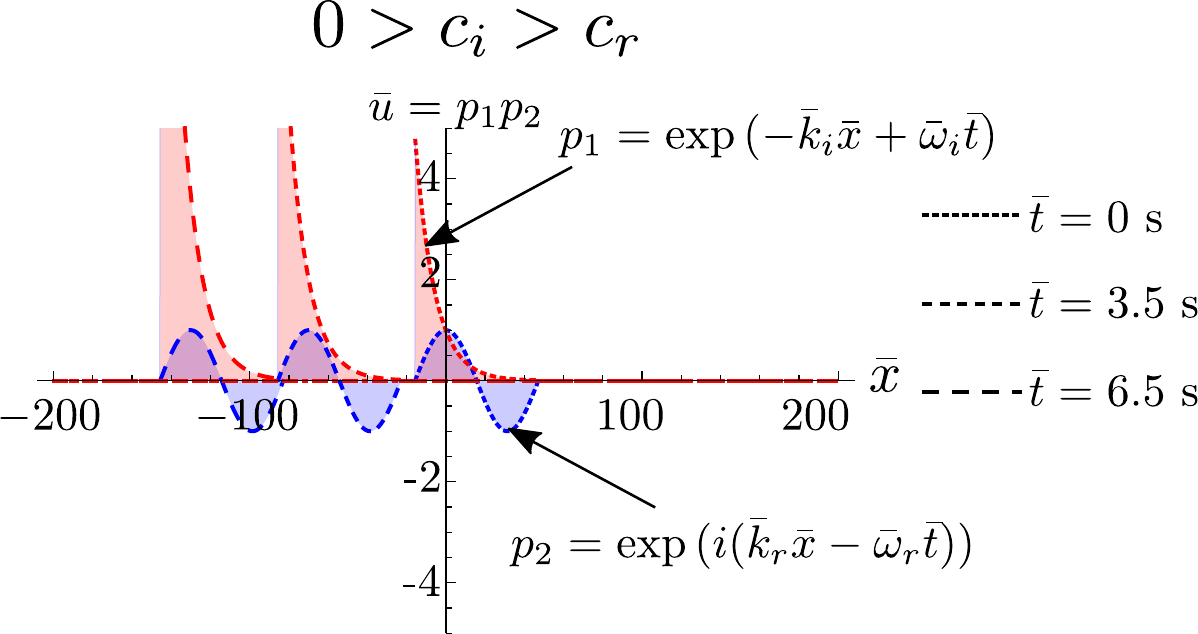}
    \end{minipage}
    \caption{Conditions for the growth of the perturbation. Left: Evolution of the perturbation (blue curve) and its amplitude (red curve) at different times for $c_i>c_r>0$. Right:Evolution of the perturbation (blue curve) and its amplitude (red curve) at different times for $0>c_i>c_r$. The propagating pulse is the multiplication of the red and blue curves.}
     \label{fig:cr_ci_2}
\end{figure}

\begin{figure}[]
    \centering
    \begin{minipage}{.45\textwidth}
    \centering
    \includegraphics[width=0.9\textwidth]{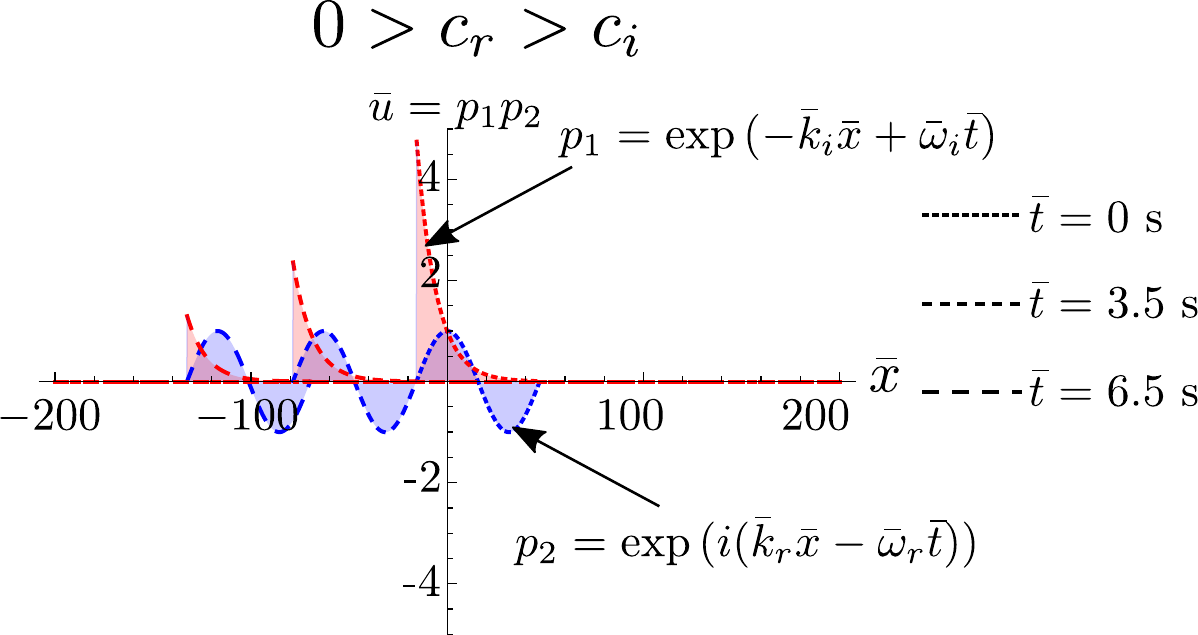}
	\end{minipage}\qquad
	\begin{minipage}{.45\textwidth}
    \centering
    \includegraphics[width=0.9\textwidth]{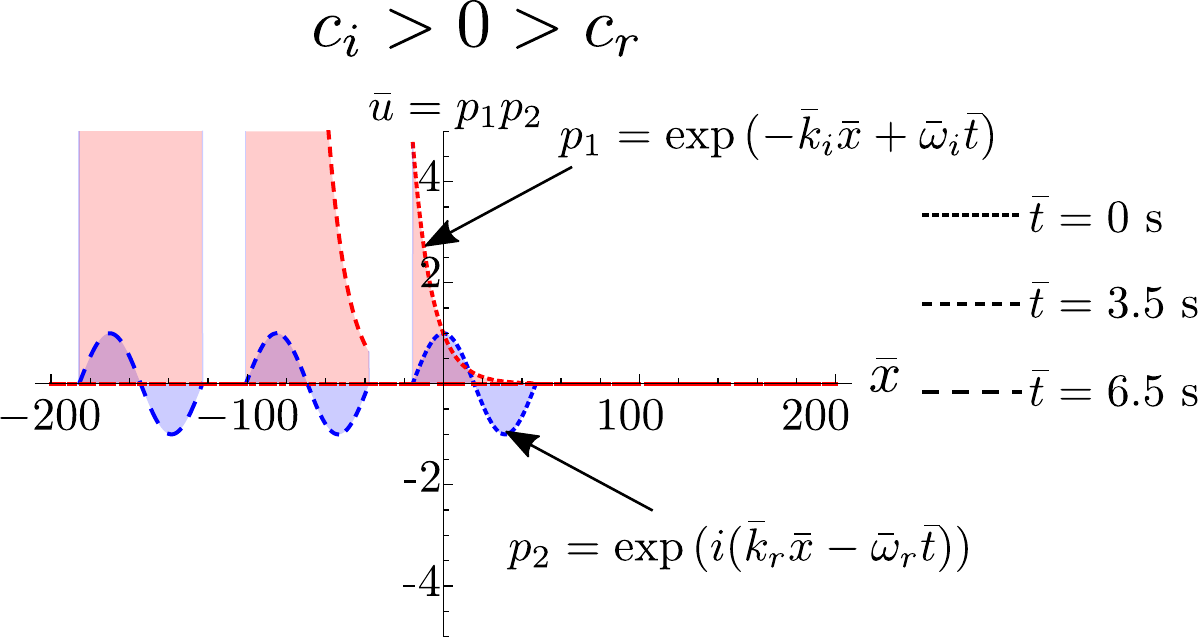}
    \end{minipage}
    \caption{Conditions for the decay and growth of the perturbation. Left: Evolution of the perturbation (blue curve) and its amplitude (red curve) at different times for $0>c_r>c_i$, decay of perturbation. Right: Evolution of the perturbation (blue curve) and its amplitude (red curve) at different times for $c_i>0>c_r$, growth of perturbation. The propagating pulse is the multiplication of the red and blue curves.}
    \label{fig:cr_ci_3}
\end{figure}

\noindent The stability of the homogeneous deformation for a strain-softening $(h<0)$ strain-rate hardening $(g>0)$ material, was examined based on the amplification (unstable) or attenuation (stable) of an arbitrary perturbation. When looking only at the case of a monochromatic propagating sinusoidal pulse, 
two velocities could be identified. These are the amplitude velocity, $c_i$, and the phase velocity, $c_f$ (see section \ref{sec: three}).
Examining equation \eqref{ubar_expanded} we notice that the amplitude term can be described as an exponential function in $\bar{x}$ traveling along $\bar{x}$-axis with velocity $c_i$. 
Similarly, the periodic part which travels with a velocity of $c_r$. A monochromatic sinusoidal pulse, whose amplitude varies with time and distance, is given as:
\begin{flalign}
&\bar{p}(\bar{x},\bar{t})=\left[H(\bar{k}_r\bar{x}-\bar{\omega}_r \bar{t})-H(\bar{k}_r\bar{x}-\bar{\omega}_r \bar{t} -2 \pi)\right]\bar{u}\exp{(-\bar{k}_i\bar{x}+\bar{\omega}_i\bar{t})}\exp{(i(\bar{k}_r\bar{x}-\bar{\omega}_r\bar{t}))},\\
&\bar{p}(\bar{x},\bar{t})=\left[H(\bar{x}-c_r\bar{t})-H(\bar{x}-c_r\bar{t} -2 \pi)\right]\bar{u}\exp{\left(-(\bar{x}-c_i\bar{t})\right)}\exp{(i(\bar{x}-c_r\bar{t}))}\label{monochromatic_pulse}.
\end{flalign} 
The Heaviside terms $H(\;\cdot\;)$ are multiplied to the original monochromatic solution to indicate the start and end of the monochromatic signal. Therefore, they travel with the velocity of the periodic wave. In this way we can describe the amplitude that corresponds to the wavelength of the pulse at a specific time. 
Based on equation \eqref{monochromatic_pulse} 
the relationship between the velocities of the two exponentials comprising the pulse is indicative of the stability and possible strain localization of the solution. In particular the following cases are possible.
\begin{itemize}
\item{$c_i<0, c_r>0$}
\item{$c_i>0, c_r>0$}
\item{$c_i>0, c_r<0$}
\item{$c_i<0, c_r<0$}
\end{itemize}
The negative signs in $c_i,c_r$ refer to the wave moving opposite to the positive direction defined by the positive $\bar{x}$-axis.
In the first case described above $c_i<0, c_r>0$ (see left part of Figure \ref{fig:cr_ci_1}), the perturbation is moving towards the positive part of the $x$-axis while the amplitude towards the negative. Due to the construction of the amplitude function (negative exponential) this has as an effect that every perturbation is attenuating with time. Therefore stability of the reference solution of homogeneous deformation is ensured and strain localization cannot take place.\\
\newline
\noindent In the second case both amplitude and phase are moving towards the positive part of the $\bar{x}$-axis as shown on the right part of Figure \ref{fig:cr_ci_1} and the left part of Figure \ref{fig:cr_ci_2}. In this case the behavior of the perturbation is defined by the relative magnitudes of the velocities $|c_i|,|c_r|$. If $|c_i|<|c_r|$ then the velocity of the negative exponential is lower than that of the perturbation. Therefore, the amplitude of the perturbation is attenuated and the reference solution is stable (see right part of Figure \ref{fig:cr_ci_1}). In the opposite case, where the perturbation travels slower than the amplitude velocity, the perturbation grows, rendering the reference solution unstable (see left part of Figure \ref{fig:cr_ci_2}). Since the amplitude is increasing the fastest at the peak behind the pulse, displacement is localizing close to the tip and localization to the smallest mesh dimension is inevitable.\\
\newline
\noindent In the third case when $c_i<0, c_r<0$, again the amplitude function and the perturbation are traveling towards the negative direction (see left part of Figure \ref{fig:cr_ci_2}, right part of Figure \ref{fig:cr_ci_3}). Again the question of stability and localization is dependent on the relative magnitudes of the two velocities $|c_i|,|c_r|$. In this case, if the perturbation is traveling slower than the amplitude $|c_r|<|c_i|$ then the amplitude of the perturbation is decreasing and no localization happens (see left part of Figure \ref{fig:cr_ci_3}). When we consider the case where $|c_r|>|c_i|$ then the amplitude of the perturbation is increasing exponentially. The amplitude is increasing the fastest for the tip closer to the front of the pulse and localization to the smallest wavelength cannot be avoided (see right part of Figure \ref{fig:cr_ci_2}.\\
\newline
\noindent In the final case $c_i>0, c_r<0$, the perturbation is moving towards the negative part of the $\bar{x}$-axis while the amplitude towards the positive (see right part of Figure \ref{fig:cr_ci_3}). Due to the negative exponential spatial profile of the amplitude function, the amplitude of the perturbation is always increasing the fastest at the tip in front of the pulse. Therefore, in this final case, the solution is unstable and  strain localization is possible with the smallest possible wavelength. 

\renewcommand{\theequation}{C.\arabic{equation}} 
\renewcommand{\thesection}{C.\arabic{section}} 
\renewcommand{\thesubsection}{C.\arabic{subsection}}   
\renewcommand\thefigure{C.\arabic{figure}}

\setcounter{figure}{0} 
\setcounter{section}{0}
\setcounter{equation}{0}  
\end{appendices}
\pagebreak
\nocite{*} 
\typeout{}

\end{document}